\documentclass[prb,preprint,showpacs,showkeys]{revtex4}
\usepackage{amssymb}
\usepackage{amsmath}
\usepackage{graphicx}
\usepackage{color}
\usepackage{soul}				
\usepackage[normalem]{ulem}
\begin{document}

\title{Pattern formation from consistent dynamical closures of uniaxial nematic liquid crystals}

\author{Humberto H\'{\i}jar\footnote{Fellow of SNI Mexico. 
E-mail for correspondence: hijar@daad-alumni.de}}
\affiliation{Facultad de Ciencias, Universidad Nacional Aut\'onoma de M\'exico, Circuito
Exterior de Ciudad Universitaria, 04510, D. F., M\'exico}
\author{Diego Marquina de Hoyos}
\affiliation{Facultad de Ciencias, Universidad Nacional Aut\'onoma de M\'exico, Circuito
Exterior de Ciudad Universitaria, 04510, D. F., M\'exico}
\author{Iv\'{a}n Santamar\'{\i}a-Holek\footnote{Fellow of SNI Mexico.
E-mail for correspondence: isholek.fc@gmail.com}}
\affiliation{UMJ-Facultad de Ciencias,  Universidad Nacional Aut\'onoma de M\'exico Campus Juriquilla, Quer\'etaro 76230, M\'exico}

\begin{abstract}
Pattern formation in uniaxial polymeric liquid crystals is studied for different dynamic closure approximations.
Using the principles of mesoscopic non-equilibrium thermodynamics in a mean-field approach, we derive a 
Fokker-Planck equation for the single-particle non-homogeneous distribution function of particle orientations and 
the evolution equations for the second and fourth order orientational tensor parameters. Afterwards, two dynamic 
closure approximations are discussed, one of them considering the relaxation of the fourth order orientational 
parameter and leading to a novel expression for the free-energy like function in terms of the scalar order 
parameter. Considering the evolution equation of the density of the system and values of the interaction parameter
for which isotropic and nematic phases coexist, our analysis predicts that patterns and traveling waves
can be produced in lyotropic uniaxial nematics even in the absence of external driving.
\end{abstract}
\maketitle

\section{Introduction}

The study of anisotropic fluids is a very active field of research in soft condensed matter 
\cite{doi001,korgerbook,degennes001,onsager001,maier001,degennes002,hess001,e-hess,hinch001,thirumalai001,larson001,wang001,
chaubal001,ilg001,hernandez,olegario1,KAC,IHK,kroeger001} in which important results have been achieved very recently. \cite{KAC,IHK} 
One of the main aspects of this study is the obtention of free energies able to describe the transition from orientational disordered to ordered 
states in relation to density or temperature conditions.\cite{doi001,korgerbook}  The evaluation of this free-energy is a difficult task\cite{hess001} 
in which mean field theories of anisotropic molecules in suspensions or pure systems have proven to be very powerful. 
The two general approaches to the determination of the free-energy function can be categorized as equilibrium and dynamic ones,
and lead to different nontrivial structural and dynamic properties of nematic liquid crystals.\cite{doi001,korgerbook,onsager001,hess001,KAC,IHK} 

Equilibrium approaches are focused on the direct calculation of the canonical distribution function, the partition function and the free-energy by taking 
into account  the general symmetry properties of the system.\cite{bingham, e-hess,ilg001,IHK} For uniaxial systems, this implies that the fourth order 
scalar orientational parameter
can be expressed in terms of the second order scalar orientational parameter. \cite{korgerbook} Recently, a step forward in the equilibrium approach
has been given in Ref. [19] where, by using the maximum entropy principle based on Gibbs entropy postulate,\cite{e-hess} a very general method for 
calculating the free-energy was proposed. In contrast, dynamic approaches to the problem are based on a closure of the evolution equations for 
the hierarchy of moments of the non-equilibrium distribution function, typically obtained from a Fokker-Planck 
equation.\cite{doi001,hess001,korgerbook,KAC,muschik} An excellent analysis of the physical consistency of several dynamic closures proposed 
in the literature is given in Ref. [18]. 

We consider that these two approaches to the problem are complementary since the equilibrium canonical distribution containing 
the specific interaction model for the system may be used to calculate both, the equilibrium free energy and the corresponding Fokker-Planck equation for the 
non-equilibrium case.\cite{doi001,hess001,korgerbook} 

In general, the results obtained by these two approaches are different even if the same interaction model is assumed. To take an example, 
for the Maier-Saupe interaction model in the uniaxial case, which introduces a parameter $U$ measuring the degree of coupling of one molecule 
with the average system surrounding it, the equilibrium approach leads to an expression of the free energy in which the dependence on $U$ enters
through one term (see, for instance, Eq. (20) of Ref. [19]). If entropic effects are adequately taken into account, other terms of this free energy appear
which are independent of the strength of interaction.\cite{IHK} 
In contrast, for this model the dynamic approach leads to expressions of a function playing the role of a free energy in which the 
parameter $U$ multiplies several terms that cannot be reduced to that of the equilibrium approach (see, for instance, Eqs. (36) and (37) of Ref. [18], 
among others). In fact, the expression of this function depends on the way in which the hierarchy of equations is closed. This situation is usual in 
non-equilibrium statistical mechanics when dealing with closure problems.  Different approximations
may have advantages under different physical conditions.

We may attribute these differences between free energies derived from equilibrium and dynamic approaches to the fact that 
dynamic closures may incorporate the effect of the relaxation of, in principle,  all the moments of the distribution function. As a consequence
of this, it is important to check the performance of the obtained results by following general criteria, such as those proposed in Ref. [18]. 
In addition, because the
equilibrium approach is general, it seems appropriate to establish a distinction by referring to a \emph{free energy} when it is derived from 
equilibrium schemes and to a \emph{free-energy like} function when the corresponding quantity is derived from dynamic calculations.\cite{doi001,KAC}



Here, we concentrate our analysis on the dynamic closure approach for uniaxial systems and its coupling with the evolution
equation of system's mass density. This non-homogenous problem is less studied in the literature despite its great interest 
because it allows to analyze the physical conditions in which an appropriate perturbation may produce patterns and/or traveling 
structures in the system. A question of great technological interest.

The existence of these spatial and temporal structures is analyzed by first using, in Section II, the Gibbs entropy postulate\cite{IHK,revMNET} 
and the rules of mesoscopic non-equilibrium thermodynamics\cite{revMNET,mnetPolim,mendez001,hijar001} in order to derive a Fokker-Planck 
equation governing the evolution of the non-homogeneous distribution of particle orientations. In section III, this equation is used in turn to 
derive the evolution equations of the second and fourth order orientational tensor parameters in the homogeneous case. Then, the dynamic  
closure problems addressed in Section IV, where we consider a recent approach\cite{KAC} to determine the forth order scalar parameter as a 
function of the second order one and then obtain the corresponding evolution equation for the scalar second order parameter. We also consider
an approximation of the dynamic equations for both scalar orientational order parameters that allows to propose a novel closure
relation leading to a novel expression for the free-energy like function. We test this closure in comparison
with those proposed in Refs. [18]  and [19] by analyzing its performance under equilibrium and flow conditions. 
In Section V we derive the non-homogeneous evolution equation for system's density and coupling parameter, and for the scalar second 
order parameter. Using the mentioned closures, we make a linear stability analysis and search for the conditions in which patterns and traveling 
structures appear.  Finally, the main conclusions are presented in Section~VI.

\section{Derivation of the orientational Fokker-Planck Equation}

We shall consider a system of $N$ elongated uniaxial molecules in a thermal bath. 
The state of each molecule is specified by the position of its center of mass, 
$\vec{r}$, and by the unitary orientational degree of freedom, $\vec{u}$, 
associated with its long axis.  The dynamics of the system can be described in terms of the 
single-particle distribution function $f\left( \vec{r}, \vec{u}, t \right)$ that is a conserved quantity and 
evolves in time by following the continuity equation
\begin{equation}
\frac{\partial f}{\partial t}=-\nabla _{i}(fV_{i})-\hat{R}_{i}\left( f\Omega
_{i}\right) ,  \label{mnet001}
\end{equation}
where, $V_{i}$ and $\Omega _{i}$ are the conjugate velocities to $x_{i}$ and 
$u_{i}$, respectively. Here, $\nabla _{i}$ is the $i$-th spatial derivative and 
$\hat{R}_{i}$ is the $i$-th component of the rotational operator
\begin{equation*}
\hat{R}_{i}=\varepsilon _{ijk}u_{j}\frac{\partial }{\partial u_{k}}.
\end{equation*}
Here, $\varepsilon _{ijk}$ represents the Levi-Civitta
antisymmetric symbol and summation over repeated indexes will be implicit
in through out this paper.

We will adopt a mean-field approach in which interactions of a single molecule with other molecules are represented
by the mean-field potential $\mathcal{U}_{\text{mf}}$ that can be a function of both position and 
orientation vectors, i.e. $\mathcal{U}_{\text{mf}} = \mathcal{U}_{\text{mf}} \left( \vec{r}, \vec{u} \right)$. 
In addition, we will also consider an external non-homogeneous time-dependent force 
characterized by the potential $\mathcal{U}_{\text{e}}$ that acts on the ensemble of molecules and drives it out 
from equilibrium. Such a force may represent, for instance, 
the influence of an electric field. \cite{mendez001,hijar001,hernandez,olegario1,muro} 
The spatial and temporal variations of the total potential field $\mathcal{U}=\mathcal{U}_{\text{mf}}+\mathcal{U}_{\text{e}}$ 
are assumed to evolve in space and time-scales large as compared to the 
time-scales of the relaxation of fluctuations. Accordingly, we assume 
the existence of a local-equilibrium probability distribution, 
$f^{\text{leq}}\left( \vec{r}, \vec{u};t\right)$, which is defined in terms of 
the previous potential by the canonical relation
\begin{equation}
f^{\text{leq}}\left( \vec{r},\vec{u},t\right) =\frac{1}{z^{\text{leq}}
\left(T;t\right) }
\exp \left[ -\frac{\mathcal{U} \left(\vec{r}, \vec{u}, t \right) }
                  {k_{B}T}\right] ,
\label{mnet002}
\end{equation}
where $z^{\text{leq}}$ is the corresponding partition function, $T$ is the
temperature and $k_{B}$ is the Boltzmann constant. The particular expression
for $f^{\text{leq}}$ can be obtained from equilibrium calculations.\cite{IHK,bingham,e-hess} 

The explicit form of Eq.~(\ref{mnet001}) in terms of 
$f\left( \vec{r},\vec{u},t\right)$ can be obtained by calculating the rate of entropy production 
of the system during its evolution in time with the help of the Gibbs entropy 
postulate~\cite{IHK,revMNET,hijar001,ivanjcp}
\begin{equation}
\mathcal{S}=-k_{B}\iint d\vec{r}d\vec{u}\ f\left( \vec{r},\vec{u},t\right) \ln \left[ 
\frac{f\left( \vec{r},\vec{u},t\right) }
{f^{\text{leq}} \left( \vec{r},\vec{u};t\right) }\right] 
+\mathcal{S}^{\text{leq}},  \label{mnet003}
\end{equation}
where $\mathcal{S}^{\text{leq}}$ is the entropy of the local-equilibrium state.  
The entropy production rate  $\sigma =\partial \left( \mathcal{S}-\mathcal{S}^{\text{leq}}\right) /\partial t$,
is therefore obtained by calculating the time derivative of Eq.~(\ref{mnet002}), 
and using Eq.~(\ref{mnet001}). This 
procedure yields 
\begin{equation}
\sigma =\frac{1}{T}\iint d\vec{r}d\vec{u}\ 
       \left[ 
       \mu \nabla _{i}\left( f\ V_{i}\right) 
      +\mu \hat{R}_{i}\left( f\ \Omega _{i}\right) 
       \right]
      +k_{B} \iint d\vec{r}d\vec{u}\ f\ 
      \frac{\partial }{\partial t}\ln f^{\text{leq}},  
\label{mnet004}
\end{equation}
where the non-equilibrium chemical potential, $\mu =\mu \left( \vec{r},\vec{u},t\right) $,
has been defined by
\begin{equation}
\Delta \mu \left( \vec{r},\vec{u},t\right) = 
k_{B}T\ln \left[ 
         \frac{f\left( \vec{r}, \vec{u},t\right)    }
              {f^{\text{leq}}\left( \vec{r},\vec{u},t\right)}
          \right] .
\label{mnet005}
\end{equation}

As we mentioned before, changes in time of the local equilibrium distribution can be neglected 
in the limit of slow external perturbations and, consequently, the second 
term on the right hand side of Eq.~(\ref{mnet004}) vanishes. In this case, Eq.~(\ref{mnet004}) 
takes the form
\begin{equation}
\sigma =-\frac{1}{T}\iint d\vec{r}d\vec{u}\ 
       \left( f V_{i}\nabla _{i}\mu
             +f \Omega _{i}\hat{R}_{i}\mu 
       \right) ,  
\label{mnet005a}
\end{equation}
where we have performed an integration by parts and assumed an infinite 
system with vanishing probability density at the boundaries in $\vec{r}$-space 
as well as periodic conditions in $\vec{u}$-space.

The entropy production rate given by Eq.~(\ref{mnet005a}) has the form of a sum 
of products of generalized currents, $V_{i}$ and $\Omega _{i}$, with
generalized forces, $\nabla_{i}\mu $ and $\hat{R}_{i}\mu $. According to the rules of mesoscopic
non-equilibrium thermodynamics, we may follow a linear response scheme in which 
currents are proportional to  forces.~\cite{revMNET,kubo001} Thus, the currents $V_{i}$ and $\Omega _{i}$ are given 
in terms of $\nabla _{i}\mu $ and $\hat{R}_{i}\mu $ by the relations
\begin{equation}
V_{i}       = - K_{ij} \nabla _{j} \mu - M_{ij}\hat{R}_{j} \mu ,  \label{mnet006}
\end{equation}
\begin{equation}
\Omega _{i} = - L_{ij} \hat{R}_{j} \mu - \tilde{M}_{ij} \nabla _{j} \mu,  \label{mnet007}
\end{equation}
where $K_{ij}$, $L_{ij} $ and $M_{ij}$ are Onsager coefficients with
$M_{ij}=-\tilde{M}_{ij}$. 
Replacing Eqs.~(\ref{mnet006}) and (\ref{mnet007}) 
into Eq.~(\ref{mnet001}), using the expressions for the local equilibrium 
distribution~(\ref{mnet002}) and the non-equilibrium chemical
potential~(\ref{mnet005}), we finally obtain a closed differential
equation for $f$:
\begin{equation}
\frac{\partial f}{\partial t} =
\nabla _{i} \left[ D_{ij} \left( \nabla _{j} f
                 + \frac{f}{k_{B}T} \nabla _{j} \mathcal{U} 
                          \right)
            \right] 
+
\hat{R}_{i} \left[ \mathcal{D}_{ij} \left( \hat{R}_{j} f 
                 + \frac{f}{k_{B}T} \hat{R}_{j} \mathcal{U}  
                                    \right) 
            \right] .
\label{mnet008}
\end{equation}
Here, we have introduced the diffusion tensors 
$D_{ij} = k_{B}T K_{ij}$ and $\mathcal{D} _{ij} = k_{B}T L_{ij}$. For anisotropic molecules both rotational and translational diffusion 
tensors depend on the orientational degrees of freedom $u_{i}$. The translational diffusion tensor can be expressed in terms of the 
parallel $D_{\parallel}$ and perpendicular $D_{\perp}$ (to the symmetry axis) coefficients in the form~\cite{korgerbook} 
\begin{equation}
D_{ij} = D_{\parallel} u_{i} u_{j}
        +D_{\perp} \left( \delta_{ij} - u_{i} u_{j} \right) 
       = \bar{D} \delta_{ij} 
       + D_{a} \left( u_{i} u_{j} - \frac{1}{3}\delta_{ij} \right),
\label{mnet009}
\end{equation} 
where we have defined $\bar{D} = \left( D_{\parallel} +2 D_{\perp} \right) /3$ and
$D_{a} = D_{\parallel} - D_{\perp}$.

At mesoscopic level, the FPE Eq.~(\ref{mnet008}) provides the complete dynamical description of 
an ensemble of uniaxial molecules subjected to mean-field and external forces.  It coincides with the ones derived in Refs.~[1] and [14] 
and therefore shows their compatibility with the second law of thermodynamics. A detailed analysis of the relation of this 
equation with its counterpart derived from kinetic theory arguments in different physical situations can be found in Refs. [2] and [21]. In addition 
it is convenient to mention that the present formalism can be generalized to consider the effect of the relaxation of the fluctuating velocities 
(angular and translational) of the molecules in the macroscopic relaxation of the system. This may lead to non-trivial diffusion effects, 
especially in the case of flowing systems.\cite{PRE2001,PRE2009}

\section{Hierarchy of macroscopic dynamic equations}

Experimental data characterizing the behavior of a system is more frequently obtained through the time course of 
the components of the tensor order parameter and higher order 
moments of the distribution function $f$, than from the form and behavior of the
distribution itself. Thus, it is convenient to derive the evolution equations for these 
moments. For simplicity, this will be done for the case when non-homogeneities
can be neglected and therefore we shall assume that no external fields are applied 
and that the description can be carried out in terms of the reduced probability 
density
\begin{equation}
g\left(\vec{u},t\right) = \int d\vec{r} f\left(\vec{r},\vec{u},t\right),
\label{hierarchy001}
\end{equation}
which obeys the reduced FPE
\begin{equation}
\frac{\partial g}{\partial t} = 
\hat{R}_{i} \left[ \mathcal{D}_{ij} \left( \hat{R}_{j} g
                 + \frac{g}{k_{B}T} \hat{R}_{j} \mathcal{U}
                                    \right)
            \right] .
\label{hierarchy001a}
\end{equation}
Here, we will assume that each molecule evolves in the presence of a 
Maier-Saupe mean-field potential of the form\cite{korgerbook,doi001} 
\begin{equation}
\mathcal{U}_{\text{mf}} = -\frac{3}{2} U k_{B} T u_{i} u_{j} S_{ij},
\label{hierarchy002}
\end{equation}
where $S_{ij}$ is the second order orientational tensor whose mathematical 
definition is given in  Eq. (\ref{hierarchy003}) below. In addition, $U$ is a parameter that measures 
the degree of coupling of the particle with its surroundings
and depends on the molecular structure and interactions.\cite{hill001}   For a concentrated solution 
of rigid rodlike polymers of length $L$ and diameter $b$, $U$ is proportional to $\rho bL$, where 
$\rho$ is the number of polymers per unit volume.~\cite{doi001}

Following Ref. [1], the moments of $g$ are defined through symmetric traceless tensors of
rank two, four, etc., corresponding to a multipolar expansion of the orientational
degree of freedom. In the next, the dyad product of a vector, $\vec{u}$, with itself $n$ times, will be 
denoted by $\mathbf{u}_{(n)}$ and $\mathbf{1}$ will denote the unitary tensor
of rank two.  General tensors  will be denoted by blackboard bold characters, 
$\mathbb{A}$, $\mathbb{B}$, etc. Indices notation will be also used when convenient. 
Finally, $\left[ \mathbb{A}\right]^s$ will denote the symmetric part of the tensor $\mathbb{A}$.

The symmetric traceless tensors of rank two, four and six, are respectively defined by the 
following averages over the distribution $g$,\cite{korgerbook}
\begin{equation}
\mathbb{S} = \bigg\langle \mathbf{u}_{(2)} -\frac{1}{3} \mathbf{1}  \bigg\rangle ,
\label{hierarchy003}
\end{equation}
\begin{equation}
\mathbb{W} = \bigg\langle \mathbf{u}_{(4)} 
            -\frac{6}{7}  \ \left[ \mathbf{u}_{(2)} \mathbf{1} \right]^s
            +\frac{3}{35}      \left[\mathbf{1}       \mathbf{1}     \right]^s
             \bigg\rangle ,
\label{hierarchy004}
\end{equation}
\begin{equation}
\mathbb{Z} = \bigg\langle \mathbf{u}_{(6)} 
            -\frac{15}{11} \left[ \mathbf{u}_{(4)} \mathbf{1}            \right]^s
            +\frac{15}{33}  \left[ \mathbf{u}_{(2)} \mathbf{1} \mathbf{1}  \right]^s
            -\frac{5}{231}      \left[ \mathbf{1}       \mathbf{1} \mathbf{1}  \right]^s
             \bigg\rangle.
\label{hierarchy005}
\end{equation}

Following an usual scheme in studying the mesoscopic dynamics of nematic liquid crystals,~\cite{doi001} we shall 
assume that rotational diffusion is isotropic, that is, $\mathcal{D}_{ij} = \mathcal{D} \delta_{ij}$ 
with $\mathcal{D}$ a constant. This assumption is consistent with the preaveraging approximation 
where a constant preaveraged rotational diffusion coefficient is introduced.\cite{doi001,kroeger001,larson-otti1991}

Thus, using the reduced FPE  Eq.~(\ref{hierarchy001a}), and the previous definitions,
the explicit evolution equations obtained for $\mathbb{S}$ and $\mathbb{W}$  are
\begin{equation}
\frac{\partial}{\partial t} \mathbb{S} = -6\mathcal{D} \mathbb{S}
            + 6\mathcal{D} U \left[ 
              \frac{1}{5} \mathbb{S} 
            + \frac{3}{7} \left( \mathbb{S} \cdot \mathbb{S} 
                          -\frac{1}{3} \mathbb{S} :  \mathbb{S} \, \mathbf{1} \right)
            - \mathbb{W} : \mathbb{S}
            \right] ,
\label{hierarchy006}
\end{equation}
\begin{eqnarray}
\frac{\partial}{\partial t} \mathbb{W} & = & -20 \mathcal{D} \mathbb{W}
            + \frac{4}{49} \mathcal{D} U 
              \left(
              35 \left[ \mathbb{S} \mathbb{S} \right]^s + 2 \mathbb{S} :  \mathbb{S} \left[ \mathbf{1} \mathbf{1} \right]^s
             -20 \left[ \mathbb{S} \cdot \mathbb{S} \, \mathbf{1}  \right]^s
            \right) \nonumber \\
            & & 
            + \frac{36}{11} \mathcal{D} U
            \left(
              \left[ \mathbb{S} \cdot \mathbb{W} \right]^s - \frac{3}{7} \left[ \mathbb{S} : \mathbb{W} \, \mathbf{1} \right]^s
            \right)
            -12 \mathcal{D} U \mathbb{Z} : \mathbb{S} .
\label{hierarchy007}
\end{eqnarray}

Equations (\ref{hierarchy006}) and (\ref{hierarchy007}) are the first two of 
an infinite hierarchy of coupled differential equations for the moments of the 
orientational distribution. It is important to notice that, in general, the evolution 
equation for the moment of order $n$ has terms containing the moments 
of order $n-2$ and $n+2$, thus coupling the whole hierarchy of equations. This is a 
well known result that is explained in whole detail in the classical textbook by Kr\"oger
where the complete hierarchy is explicitly calculated.~\cite{korgerbook}

An important fact to mention is that the characteristic relaxation time for each one of 
the equations in this hierarchy is proportional to $\mathcal{D}^{-1}$. As a consequence of this, it is not possible
to give a general criterium based on time scales to cut the hierarchy of equations and search for consistent 
dynamic closures. Due to this fact, many dynamic closures to the problem have been 
proposed in the literature to cut the hierarchy. The validity and physical consistency of the most representative 
of them has been analyzed in detail in Ref. [18].
However, for comparison purposes, let us recall the classical decoupling approximation proposed in Refs.~[1, 2, 7],
where
\begin{equation}
\langle \mathbf{u}_{(4)} \rangle : \mathbb{S} 
= \langle \mathbf{u}_{(2)} \rangle \langle \mathbf{u}_{(2)} \rangle : \mathbb{S} .
\nonumber
\end{equation} 
In terms of the tensors $\mathbb{S}$ and $\mathbb{W}$, the previous relation can be
written in the form
\begin{equation}
\mathbb{W} : \mathbb{S} = -\frac{2}{15} \mathbb{S} -\frac{4}{7} \mathbb{S} \cdot \mathbb{S}
                         +  \mathbb{S} : \mathbb{S} \, \mathbb{S} 
                        + \frac{4}{21} \mathbb{S} : \mathbb{S} \, \mathbf{1} .
\label{hierarchy008}
\end{equation}

Indeed, by replacing Eq.~(\ref{hierarchy008}) into Eq.~(\ref{hierarchy006}), we
recover the well known closed expression for the order parameter of the Doi-Hess model
\begin{equation}
\frac{\partial}{\partial{t}} \mathbb{S} =
-6 \mathcal{D} \left[ \left( 1-\frac{U}{3} \right) \mathbb{S} 
                    -U\left( \mathbb{S} \cdot \mathbb{S}  
                    -\frac{1}{3} \mathbb{S} : \mathbb{S} \, \mathbf{1} \right)
                    +U \mathbb{S} : \mathbb{S} \, \mathbb{S}
              \right] .
\label{hierarchy009}
\end{equation}
This approximation, as the other ones analyzed in Ref. [18], truncates the infinite hierarchy of equations 
for the moments of the distribution to the lowest possible order. An important fact to mention here is that 
the last three terms at the right hand side of the previous equation are proportional to the coupling parameter $U$. 
As mentioned in the introduction, this is a characteristic result of the dynamic closures in uniaxial systems. The corresponding free-energy 
like function that may be obtained from it by setting the time derivative equal to zero, will contain only one term 
independent of $U$, which arises from the first term in (\ref{hierarchy009}) and is related to the entropic force term 
of the reduced Fokker-Planck equation (\ref{hierarchy001a}).\cite{doi001} 

\section{Dynamic closures for the evolution equations of the scalar order parameters}

We shall consider here a uniaxial nematic liquid crystal. For 
simplicity, it will be assumed to be initially oriented by an external 
field along the unitary vector $\vec{n}$, which is called the 
director. In the following, the dynamics of this phase will be described in
terms of the scalar order parameters $S$, $W$ and $Z$, which are respectively 
defined in terms of $\vec{n}$ and the order parameters tensors 
$\mathbb{S}$, $\mathbb{W}$ and $\mathbb{Z}$, Eqs.~(\ref{hierarchy003})-(\ref{hierarchy005}), 
by\cite{korgerbook}
\begin{equation}
S = \frac{3}{2} n_{i} S_{ij} n_{j} 
  = \langle P_{2} \left( \vec{u}\cdot\vec{n} \right) \rangle,
\label{closure00a}
\end{equation}
\begin{equation}
W = \frac{35}{8} n_{i} n_{j} W_{ijkl} n_{k} n_{l}
  = \langle P_{4} \left( \vec{u}\cdot\vec{n} \right) \rangle,
\label{closure00b}
\end{equation}
\begin{equation}
Z = \frac{231}{16} n_{i} n_{j} n_{k} Z_{ijklpq}  n_{l} n_{p} n_{q}
  = \langle P_{6} \left( \vec{u}\cdot\vec{n} \right) \rangle,
\label{closure00c}
\end{equation}
where $P_{m}$ is the Lengendre polynomial of order $m$. Notice that $S$, 
$W$, and $Z$ are zero in the isotropic limit and $1$ in the completely ordered phase.

In terms of $S$, $W$ and $Z$, the following closure relation can be written for $\mathbb{Z}$ in
terms of $\mathbb{S}$ and $\mathbb{W}$,
\begin{equation}
\mathbb{Z} = \frac{Z}{S W}
             \left\{
                   \left[ \mathbb{W}  \mathbb{S} \right]^{s}
     -\frac{8}{11} \left[ \mathbb{S} \cdot  \mathbb{W} \mathbf{1} \right]^{s}
     +\frac{4}{33} \left[ \mathbb{S} :  \mathbb{W} \mathbf{1 1} \right]^{s}
             \right\} ,
\label{closure00d}
\end{equation}
where the right hand side contains the symmetric traceless part of the
product of $\mathbb{W} \mathbb{S}$. This explicit relation representing 
$\mathbb{Z}$ is indeed a particular case of the general closure equation for 
moments of arbitrary order derived in Ref.~\cite{KAC}.

Using Eq.~(\ref{closure00d}) the evolution equations for the first two scalar 
moments can be obtained by projecting Eqs.~(\ref{hierarchy003}) and 
(\ref{hierarchy004}) on $\vec{n}$. The result of this procedure is 
\begin{equation}
\frac{\partial S}{\partial t}  =  -6 \mathcal{D} \left( 1 -\frac{ U }{5}\right) S
                                  + \frac{6}{7} \mathcal{D} U \, S^{2}
                                  - \frac{72}{35} \mathcal{D} U \, S\, W ,
\label{closure00e}
\end{equation}
\begin{equation}
\frac{\partial W}{\partial t}  =  -20 \mathcal{D} \, W + \frac{20}{7} \mathcal{D} U \, S^{2}
                                  + \frac{60}{77} \mathcal{D} U \,S\, W
                                  - \frac{40}{11} \mathcal{D} U \,S\, Z .
\label{closure00f}
\end{equation}

Notice that an infinite hyerarchy of coupled equations for the
scalar moments is obtained, which must be closed at a certain level. In the
subsequent sections, two different closure relations will be discussed.

\subsection{Parametric closure}

With the purpose of closing the infinite hyerarchy of coupled equations for the
scalar order parameters, approximate relationships between $S$, $W$ and $Z$ can be obtained 
based on the uniaxial orientational distribution of the Maier-Saupe type~\cite{KAC}
\begin{equation}
g \left(\vec{u}\right) = G \exp\left[ a \vec{u}\cdot\mathbb{A}\cdot\vec{u}\right],
\label{closure00g}
\end{equation}
which is a special case of the Bingham distribution.~\cite{bingham} Here, $a$ is a parameter, 
$\mathbb{A}$ is a symmetric traceless matrix and $G$ is the normalization constant.

Using Eq.~(\ref{closure00g}), $S$, $W$ and $Z$ can be calculated in terms 
of $a$, from which the numerical relation between these quantities can be obtained. 
From this analysis it has been shown in Ref.~\cite{KAC,e-hess} that $W$ and $Z$ can be very 
well approximated in terms of $S$ by 
\begin{equation}
W = S \left( 1 - \left(1- S\right)^{\nu} \right) ,
\label{closure00h}
\end{equation}
with $\nu = 3/5$; and 
\begin{equation}
Z = S^{6}, 
\label{closure00i}
\end{equation}
respectively. The first expression can be used in Eqs.~(\ref{closure00e}) and 
(\ref{closure00f}) in order to close the hyerarchy of 
equations upto order $S$, while the latter allows for closing it at order $W$.

Equation~(\ref{closure00h}) will be referred hereafter as the 
parametric closure. It has been shown to fullfill the constraints 
imposed by the nematic symmetry and to be correct in the isotropic and 
totally aligned cases. It also yields the following evolution equation for $S$
\begin{equation}
\frac{\partial S}{\partial t}  = -6\mathcal{D} \left[
            \left( 1 - \frac{U}{5} \right) S
            -\frac{U}{7} S^{2}
            + \frac{12}{35}U S^{2} (1 -\left(1- S\right)^{\nu}) \right].
\label{closure00j}
\end{equation}

Dynamic equations of the form of the previous one, have been extensively 
used in the literature of the field since the pioneer works of Doi and Hess in
order to identify a function $A = A\left( U, S\right)$ through the phenomenological
relation\cite{doi001,KAC}
\begin{equation}
\frac{\partial S}{\partial t}  = -L \frac{\partial A}{\partial S},
\label{closure00k}
\end{equation}
such that $A$ plays the role of a free energy. As we mentioned previously, 
it is convenient to refer to this quantity as a \emph{free-energy like} function since
it differs from those calculated following equilibrium techniques. In the present case, 
the dynamic closure was implemented by calculating the equilibrium averages of the second and
fourth order orientational parameters, and using an interpolation function directly in the evolution 
equation for the second order orientational parameter.\cite{KAC} In the following subsection, we will show 
how the dynamic closure approximation is modified when the relaxation of the fourth
order orientational parameter is also taken into account.

For the parametric model, $A$ has the explicit form originally obtained in Ref.~\cite{KAC}
\begin{equation}
A = \frac{1}{2} \left( 1 -\frac{U}{5}\right)S^{2} +\frac{U}{15} S^{3}
  +\frac{12 U \left(1-S^{1+\nu}\right)\left\{ 2 + S\left(1+\nu\right)\left[2+ S\left(2+\nu \right)\right] \right\} }
        {35\left(1+\nu \right)\left(2+\nu \right)\left(3+\nu\right)},
\label{closure00l}
\end{equation}
and predicts that for $U < U_{1}^{*} = 4.48$, $A$ has only one 
minimum at $S=0$, corresponding to the isotropic phase. For 
$U_{1}^{*} < U < U_{2}^{*} = 5$, $A$ has two minima, one at $S=0$ and
the other one at $S > 0$, indicating that the system can be found in the nematic 
or the isotropic phase depending on the initial value of $S$. For $U \ge U_{2}^{*}$,
$A$ has a local maximum at $S=0$, and the nematic phase is the stable one.

The equilibrium values of the order parameter can be found from the 
condition
\begin{equation}
\left. \frac{\partial A}{\partial S} \right|_{S_{\text{eq}}} = 0, 
\label{closure00m}
\end{equation}
from which it follows that $S_{\text{eq}} \left( U_{1}^{*} \right) = 0.31$.

It should be stressed that although the function $A$ given by Eq.~(\ref{closure00l})
exhibits the typical behavior used for describing the I-NPT, it is not completely
consistent as long as it is finite at $S=1$, thus allowing for $S$ to take
nonphysical values, i.e. $S > 1$ for finite energies. This problem has been successfully solved
in Ref.~\cite{IHK}, where a thermodynamic free-energy has been derived by maximizing 
the Gibbs entropy postulate (\ref{mnet003}) with the isotropic state as the reference one. The
obtained expression ensures the constraint $S < 1$ for arbitrary energies.

Concerning the high order closure relation (\ref{closure00i}), it should be
mentioned that when it is replaced into Eqs.~(\ref{closure00e})
and (\ref{closure00f}), and the resulting closed system is used to find the 
equilibrium values of the order parameter $S$ by imposing static conditions, 
i.e. $\partial S / \partial t =\partial W / \partial t  = 0$, then nonphysical 
behavior is observed in $A$ since $S_{\text{eq}}$ turns out to be a decreasing function 
of $U$ for relative large values of this quantity. Therefore, it should be 
remarked that although Eq.~(\ref{closure00i}) can be used to approximate very 
well the values of $Z$, it can not be used consistently in the dynamic equations
for $S$ and $W$ by imposing the static conditions already mentioned. 
This suggests that having expressions for higher order moments which approximate 
very well their exact values, might not be sufficient to fullfill
the requirements imposed by the dynamic equations. 

\subsection{Dynamic closure}
\begin{figure}
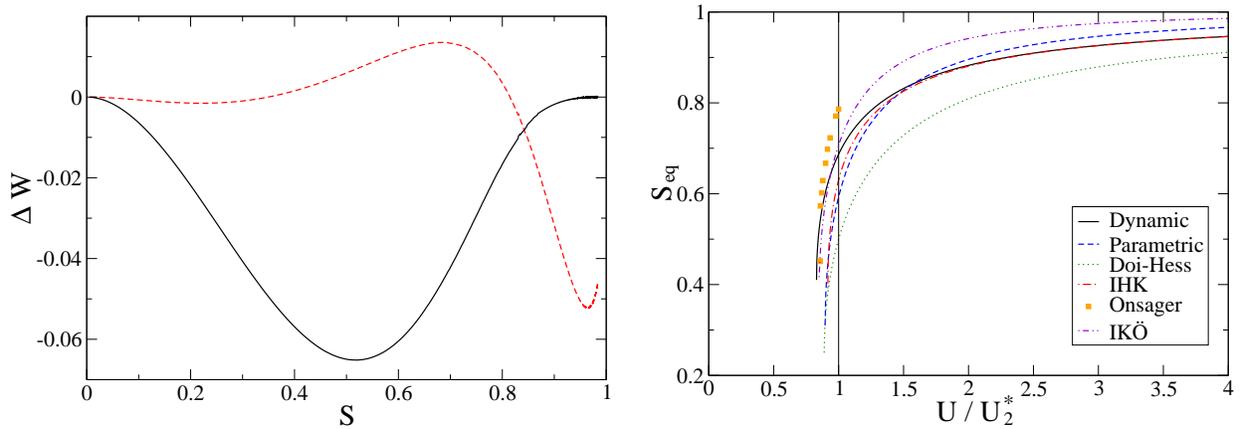

\begin{center}
\includegraphics[scale=0.32]{fig1a.eps} \,\,\,\, 
\includegraphics[scale=0.32]{fig1b.eps}\vfill
\end{center}
\caption{Left panel: Performances of the parametric closure, Eq.~(\ref{closure00h}) (dashed line),
and the dynamic closure, Eq.~(\ref{closure00n}) (solid line). $\Delta W$ 
represents the difference of the approximated expressions~(\ref{closure00h}) and 
(\ref{closure00n}), with respect to the numerical value of the order parameter $W$ 
obtained from the orientational distribution~(\ref{closure00g}). Right panel: 
Equilibrium values of the order parameter, $S_{\text{eq}}$, for the nematic phase. The black solid 
line corresponds to the dynamic closure (\ref{closure00p}), which incorporates the effects of the relaxation dynamics of $S$ and $W$. 
Orange square symbols were taken from Ref.~[14] and correspond to the numerical solution of the Onsager model. 
The red dashed-dotted line corresponds to the IHK model\cite{IHK} that incorporates entropic effects in the Maier-Saupe potential
whereas the blue dashed line comes from the (KAC)\cite{KAC} parametric closure~(\ref{closure00h}).
IK\"O model\cite{ilg001} is represented by the purple dashed double-dotted line and finally the green dotted line 
represents the classical solution obtained from Doi-Hess model.\cite{doi001}
}
\label{figure_001}
\end{figure}

Here we shall introduce an alternative closure relation for $W$ in terms of $S$,
which has the explicit dependence
\begin{equation}
W = S^{\frac{10}{3}} ,
\label{closure00n}
\end{equation} 
and will be referred hereafter as the dynamic closure relation
since it is motivated by the explicit form of the evolution 
equations for $S$ and $W$, Eqs.~(\ref{closure00e}) and (\ref{closure00f}).
Indeed, it can be verified that Eq.~(\ref{closure00n}) is exact in two
important limiting cases, namely: for $U \ll 1$, i.e. in the isotropic 
phase; and for $U \gg 1$, i.e. close to the completely aligned phase.

Consequently, this new closure relation is not intended to be exact, 
but it is an interpolating expression which is consistent with the time
evolution of the scalar order parameters. Furthermore, it turns out to 
approximate the correct parametric relation within the 
same accuracy degree than Eq.~(\ref{closure00h}). This is explicitely
shown in the left panel of Fig.~\ref{figure_001}, where the deviations from the closure
relations Eqs.~(\ref{closure00h}) and (\ref{closure00n}) from the
exact numerical parametric value of $W$ are presented. It can be noticed
that the maximum deviation observed for the parametric closure, Eq.~(\ref{closure00h}), 
is $0.052$; while the corresponding value for the dynamic approximation, 
Eq.~(\ref{closure00n}), is $0.065$. Notice also that the parametric closure 
exhibits its maximum deviations in the region $S \simeq 1$, where the dynamic 
closure approximates $W$ better.   
\begin{center}
\begin{table}
\begin{tabular}{|c|c|c|c|c|c|} \hline
Model             & Equilibrium (IHK)  & Parametric (KAC)   &  Dynamic  (HMS)  & \,\,Doi \,\,           &   \,\, IK\"O  \,\,     \\ \hline  \hline
$U_{1}^{*} $  &  $4.59$     &          4.48               &    4.15    &  $2.67$   &  $6.22$    \\ \hline
$U_{2}^{*}$   &  $5.0$       &          5.0             &     5.0   &  $3.0  $     & $7.34$   \\ \hline
$S_{\text{eq} } \left( U_{1}^{*} \right)$ &  $0.39$   &   0.31  &    0.41   & $0.25 $    & $0.37$  \\ \hline
\end{tabular}
\caption{Parameters characterizing the I-NPT in diverse models. $U_{1}^{*}$
represents the smallest value of $U$ at which the nematic phase can be 
observed, while $U_{2}^{*}$ is the strength of the mean field interaction 
at which the isotropic phase becomes unstable. $S_{\text{eq}}\left( U_{1}^{*}\right)$ 
is the equilibrium value of the order parameter at the I-NPT. This comparison
can be extended by considering Ref.~\cite{KAC}, where an extensive study which 
includes more models can be found.}
\label{table001}
\end{table}
\end{center}

The dynamic closure relation, Eq.~(\ref{closure00n}), yields the following 
dynamic equation for $S$,
\begin{equation}
\frac{\partial S}{\partial t}  = -6\mathcal{D} \left[
            \left( 1 - \frac{U}{5} \right) S
            -\frac{U}{7} S^{2}
            + \frac{12}{35}U S^{\frac{13}{3}}  \right],
\label{closure00o}
\end{equation}
and the corresponding function $A = A\left( U, S\right)$ for this approximation
reads
\begin{equation}
A =\frac{1}{2} \left( 1 - \frac{U}{5} \right) S^{2}
            -\frac{U}{21} S^{3}
            + \frac{9}{140}U S^{\frac{16}{3}} ,
\label{closure00p}
\end{equation}
which predicts the values of the parameters characterizing the I-NPT $ U_{1}^{*}$,
$ U_{2}^{*}$ and $S_{\text{eq}}\left(U_{1}^{*}\right)$,  shown in Table~\ref{table001}, 
where these parameters are compared with those obtained from other models
including the parametric closure.
\begin{figure}
{\includegraphics[width=80mm, height=40mm]{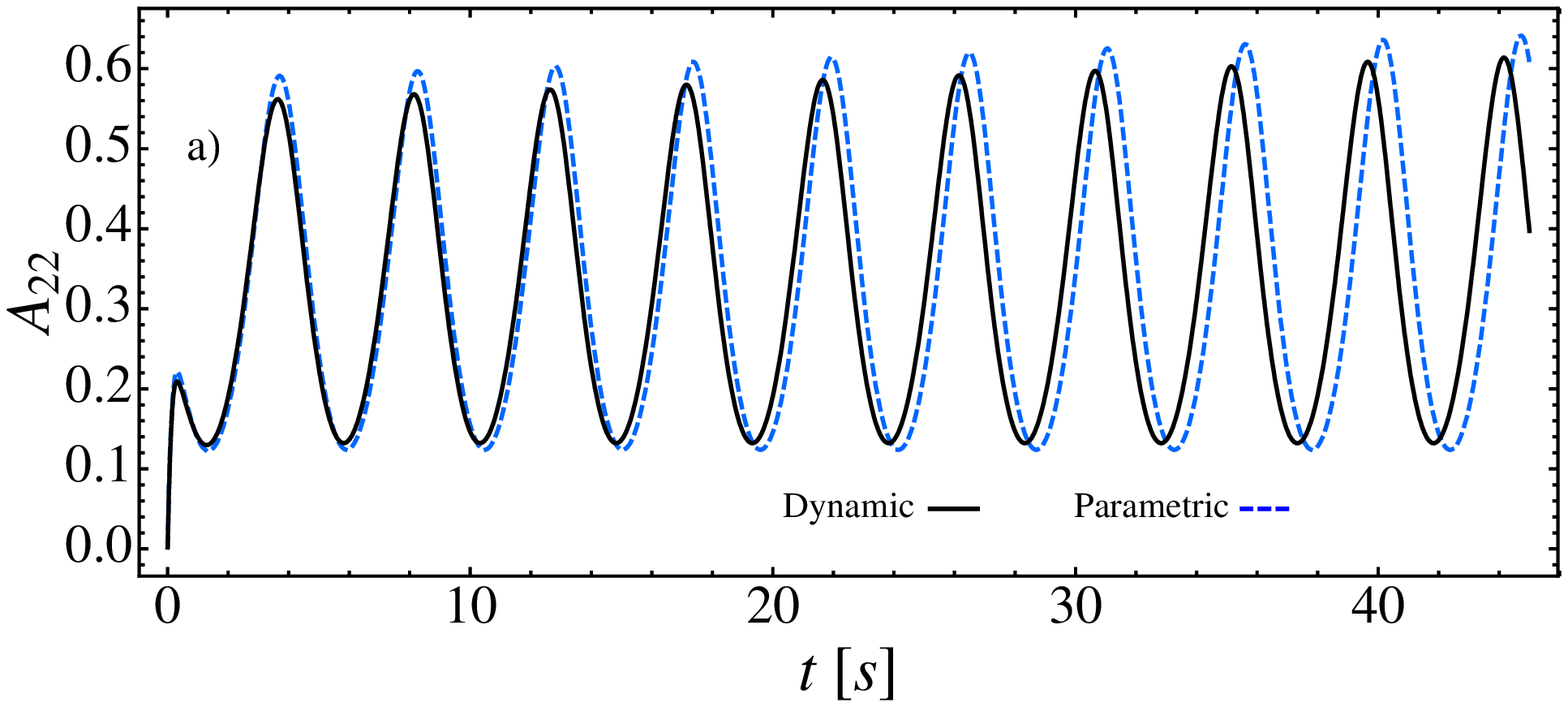}} {\includegraphics[width=80mm, height=40mm]{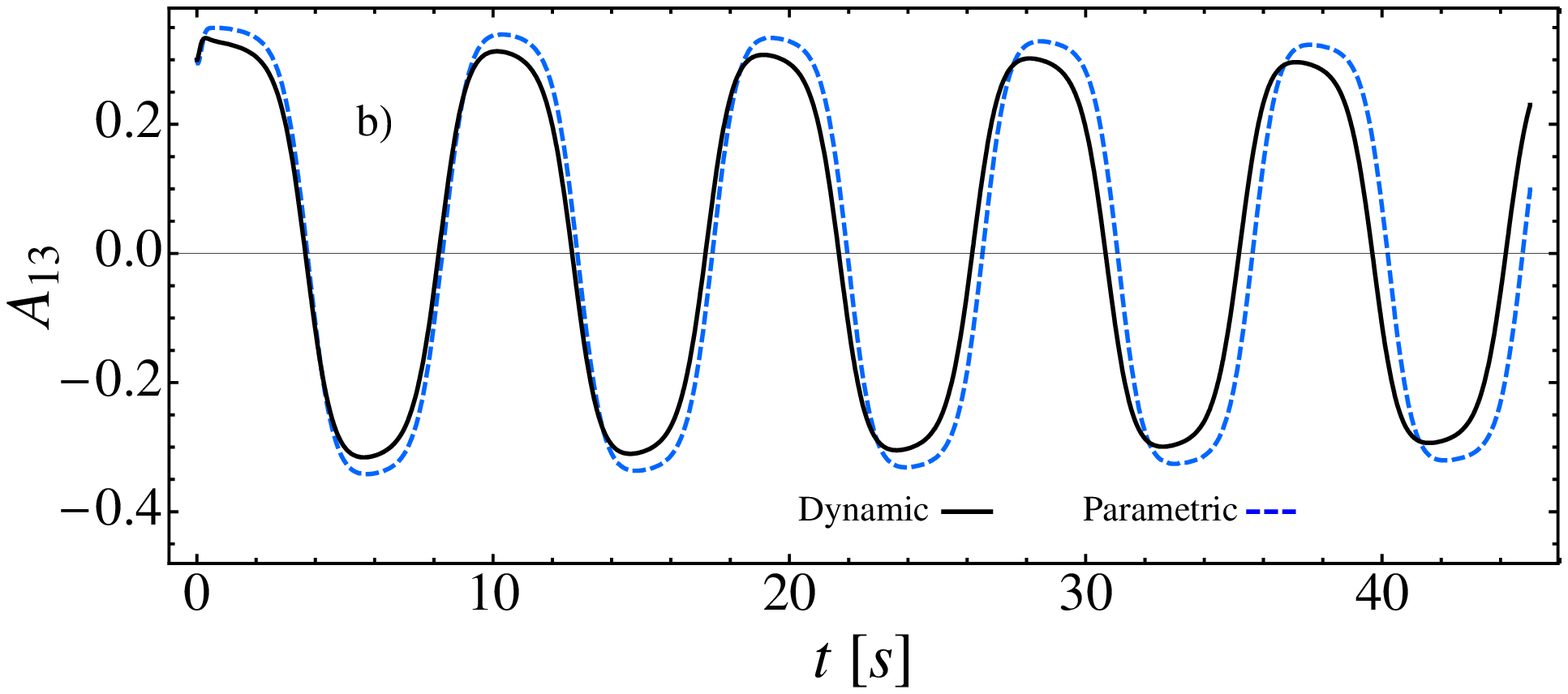}}
\caption{Oscillations of two elements of the tensor $\mathbb{A}$ in terms of time
obtained after numerically solving Eq. (\ref{flow00b}).
In a) we present the diagonal component $A_{22}$, whereas in
b) the component $A_{13}$ is shown. The values of 
the parameters used were ${\mathcal D}=0.3$, U$=9$, $Q=1.5$ and the magnitude of the
shear rate $|\nabla \vec{v}|=1.5$.}
\label{figure_002-new}
\end{figure}

\begin{figure}
\includegraphics[scale=0.40]{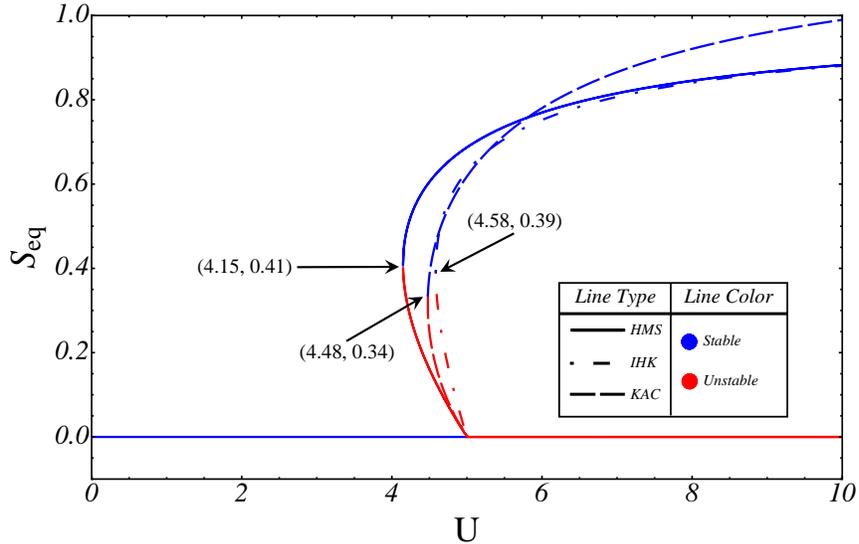} \\
\caption{Bifurcation diagram of the scalar order parameter $S_{\text{eq}}$  as a function of the interaction strength $U$, 
as obtained from the parametric and the dynamic closures, Eqs.~(\ref{closure00h}) (denoted by KAC, dashed line) and 
(\ref{closure00n}) (denoted by HMS, solid line). We also included the equilibrium closure reported in Ref. \cite{IHK}  
(denoted by IHL,  dash-dotted line).  The blue portion of the lines indicates stable states whereas the red portion unstable states. 
The purple portion indicates a stable region for KAC and HMS, and unstable for IHK. }
\label{figure_002}
\end{figure}

The equilibrium values of the scalar order parameter in the nematic phase for
the dynamic closure can be found from Eqs.~(\ref{closure00m})
and (\ref{closure00p}). The solution is shown on the right panel of Figure~\ref{figure_001} as 
function of the reduced interaction strength $U / U^{*}_{2}$. Figure~\ref{figure_001} 
also shows $S_{\text{eq}}$ for other models including the classical model of Doi and 
Hess~\cite{doi001}; the numerical solution of the Onsager excluded volume theory as it
appears in Ref.~[14]; the model proposed in Ref.~\cite{ilg001} by Ilg, Karling and \"Otinger
(IK\"O), where a generalized mean-field interaction is introduced; and the solution obtained from
the parametric KAC closure Eq.~(\ref{closure00h}). It can be observed that our 
model predicts a wider range of values of $U$ for the
coexistence of the isotropic and nematic phases than all the other models.
Finally, Figure~\ref{figure_002} shows the bifurcation diagram obtained from
the parametric and the dynamic models. 

  \subsection{Performance of the parametric and dynamic closures under flow conditions} 
In order to rigorously test the performance of the novel closure approximation
Eq.~(\ref{closure00n}), we will study its behavior under flow conditions.
For simplicity, we will restrict our analysis to a comparison with respect to
the performance of the parametric closure Eq.~(\ref{closure00h}), which in turn has 
been compared with several closure models as well as with exact results obtained
from the numerical solution of the FPE.~\cite{KAC} Our comparison is justified
since in the latter case it has shown that Eq.~(\ref{closure00h}) is valid with high 
precision for a wide range of conditions. 

We thus consider an elongated molecule immersed in a fluid with a velocity gradient 
and subject to the potential $\mathcal{U}$. The FPE corresponding to this problem 
has been calculated by considering the hydrodynamic torques which the fluid exerts 
on the molecule.~\cite{doi001,hess001} This calculation yields
\begin{equation}
\frac{\partial g}{\partial t} = 
-\omega_{i} \hat{R}_{i} g
-\frac{1}{2} P \hat{R}_{i} \left[ g \hat{R}_{i} \left( u_{j} u_{k} K_{jk} \right) \right]                               
+\hat{R}_{i} \left[ \mathcal{D}_{ij} \left( \hat{R}_{j} g
                 + \frac{g}{k_{B}T} \hat{R}_{j} \mathcal{U}
                                    \right)
            \right],
\label{flow00a}
\end{equation}
where $\omega_{i} = \frac{1}{2} \varepsilon_{ijk} \nabla_{j} v_{k}$ is the
vorticity, $K_{ij} = \frac{1}{2} \left(\nabla_{i} v_{j} + \nabla_{j} v_{i} \right)$
is the symmetric part of the velocity gradient, and
$P = (Q^{2}-1)/(Q^{2}+1)$ is the shape factor of an ellipsoid of revolution
with axis ratio $Q$. 

Once again, we will consider only a mean field potential of the Maier-Saupe
type, Eq.~(\ref{hierarchy002}), and calculate the evolution equation for the 
second order tensor, 
$\mathbb{A} = \langle \mathbf{u}_{(2)}\rangle = \mathbb{S} + \frac{1}{3} \mathbf{1}$, 
in the presence of flow directly from Eq.~(\ref{flow00a}). This equation
involves the fourth order moment $\mathbb{B} =\langle \mathbf{u}_{(4)} \rangle$, and
explicitly reads
\begin{eqnarray}
\frac{\partial }{\partial t} \mathbb{A} + \vec{v} \cdot \nabla \mathbb{A} & = &
\mathbb{A} \cdot \mathbb{L} - \mathbb{L} \cdot \mathbb{A} 
+ P \left(\mathbb{A} \cdot \mathbb{K} + \mathbb{K} \cdot \mathbb{A}  \right)
- 2 P \mathbb{K} : \mathbb{B}  \nonumber \\
& & - 6 \mathcal{D} \left( \mathbb{A} - \frac{1}{3} \mathbf{1} \right)
+ 6 \mathcal{D} U \left(  \mathbb{A} \cdot  \mathbb{A} -  \mathbb{A} : \mathbb{B} \right),
\label{flow00b}
\end{eqnarray}
where $L_{ij} = \frac{1}{2} \left(\nabla_{i} v_{j} - \nabla_{j} v_{i} \right)$
is the antisymmetric part of the velocity gradient.

This equation can be closed by using the consistent closure relationship 
between $\mathbb{A}$ and $\mathbb{B}$~\cite{KAC}
\begin{equation}
\mathbb{B} =  \alpha \left[\mathbb{A} \mathbb{A} \right]^{s} 
           -2 \beta  \left[\mathbb{A} \mathbf{1} \right]^{s}
           -2 \gamma \left[\mathbf{1} \mathbf{1} \right]^{s},
\label{flow00c}
\end{equation} 
where the coefficients $\alpha$, $\beta$ and $\gamma$ depend on
the scalar order parameters $S$ and $W$ through
\begin{equation}
\alpha = \frac{W}{S^{2}},
\label{flow00d}
\end{equation}
\begin{equation}
\beta = \frac{\alpha}{3} +\frac{2 W}{21 S} -\frac{3}{7},
\label{flow00e}
\end{equation}
\begin{equation}
\gamma = \frac{3}{70} + \frac{2 W}{45} -\frac{\alpha}{18} -\frac{2 W}{63 S}.
\label{flow00f}
\end{equation}

Thus, we replace Eq.~(\ref{flow00c}) into Eq.~(\ref{flow00b}) and solve the
result numerically for the components $A_{ij}$. We consider both, the parametric  
and the dynamic closure approximations, given by Eqs.~(\ref{closure00h}) and 
(\ref{closure00n}), respectively, in order to compare their performances. 
For simplicity, we assume a homogeneous aligment tensor $\mathbb{A}$, and 
restrict our analysis to situations similar to those reported in the recent 
literature. Figure~\ref{figure_002-new} summarizes our results. There
we present the behavior of two different components of $\mathbb{A}$, obtained
for diverse values of the parameters $P$, $\mathcal{D}$, $U$ and
$\nabla \vec{v}$ as indicated in the caption. It can be observed that the dynamic closure 
relation, Eq.~(\ref{closure00n}), performs quantitatively very well when compared
with the parametric closure Eq.~(\ref{closure00h}), and consequently can be also used 
to approximate the exact solution with high precision 
at least for the range of values considered here.

\section{Pattern formation as a consequence of the coupling of the density and the scalar order parameter}

The Fokker-Planck equation derived in Section II and the closure approximations discussed in the previous section can
be used to show that  patterns and traveling waves  may emerge for these systems by following a mechanism different to
those discussed previously in the literature, where the effect of an external driving was considered.\cite{larson1992,forest1999,forest2004} 
These non-equilibrium structures may occur when the parameter $U$ takes values in the range $U^*_1<U<U^*_2$, that 
is, when coexistence of isotropic and nematic phases is possible. 

Essentially, the existence of patterns comes from the fact that the degree of coupling $U$ may depend on the 
number density of molecules $\rho$. For a lyotropic liquid crystal we may write $U= \left({\rho}/{\rho^*}\right)U_0$, 
with $U_0$ and $\rho^*$ the characteristic energy and density of the system, respectively.~\cite{doi001} This dependence 
indicates that, when increasing the density of the system one also increases the interaction energy and 
the orientational order. This is the basic mechanism leading to the INP-T. The
formation of stationary and dynamic patterns comes from the fact that the resulting equations 
for $S(\vec{r},t)$ and $\rho(\vec{r},t)$, or similarly $U(\vec{r},t)$, constitute a set of two coupled 
equations of the reaction-diffusion type.\cite{prigo}

To show this, let us first obtain a dynamic equation for the coupling parameter $U(\vec{r},t)$, introduced in 
the description through the Maier-Saupe potential, Eq.~(\ref{hierarchy002}). We start by deriving the dynamic 
equation for the reduced probability density $\rho$ defined by
\begin{equation}\label{pattern01}
\rho(\vec{r},t) \equiv  \int f(\vec{r},\vec{u},t) d\vec{u}. 
\end{equation}
Integrating the Fokker-Planck equation (\ref{mnet008}) over the solid angle yields the following evolution 
equation for the number density
\begin{equation}\label{pattern02}
 \frac{\partial \rho}{ \partial t} = \bar{D} \nabla^2 \rho -\frac{1}{2}\bar{D} \nabla^2(US^2)- \frac{1}{2}\bar{D} \nabla \cdot \left[ S^2 \nabla U(\rho) \right],
\end{equation} 
where we have used Eq.~(\ref{hierarchy002}) and $\bar{D}$ has been assumed 
to be constant. Using the relation $\rho= \left({U}/{U_0}\right)\rho^*$ with $U_0=5$
in the left hand side of Eq.~(\ref{pattern02}) and rearranging terms we finally obtain 
\begin{equation}\label{pattern03}
 \frac{\partial U}{ \partial t} = \bar{D} \left(1-\frac{5}{\rho^*} S^2\right)\nabla^2 U - 
\frac{15}{\rho^*} \bar{D} S \left[\nabla S \cdot  \nabla U \right]-\frac{5}{2\rho^*}\bar{D}U\nabla^2S^2. 
\end{equation} 
This equation introduces the effective 
diffusion coefficient $D_{\text{eff}} = \bar{D} \left(1-\frac{5}{\rho^*} S^2\right)$. It is interesting to notice that the correcting factor
implies that when orientational order increases in the system the diffusion of the particles decreases. However, it is worth stress that molecular 
dynamics simulations predicted that the average diffusion coefficient $ \bar{D}$ may increase under this conditions.\cite{frenkel} 
This competing interplay may lead to an interesting non-trivial (non-monotonic) behavior of the effective diffusion coefficient.

To complete the description, we may use Eq.~(\ref{mnet008}) to derive a general evolution 
equation for the non-homogeneous order parameter tensor $S_{ij}\left( \vec{r} , t \right)$ 
defined in Eq. (\ref{hierarchy003}). Multiplying Eq.~(\ref{mnet008}) by $u_{i} u_{j} -\frac{1}{3} \delta_{ij}$ 
and taking the orientational average of the result, an integration by parts and some algebra yield the equation
\begin{figure}
{\includegraphics[width=75mm, height=30mm]{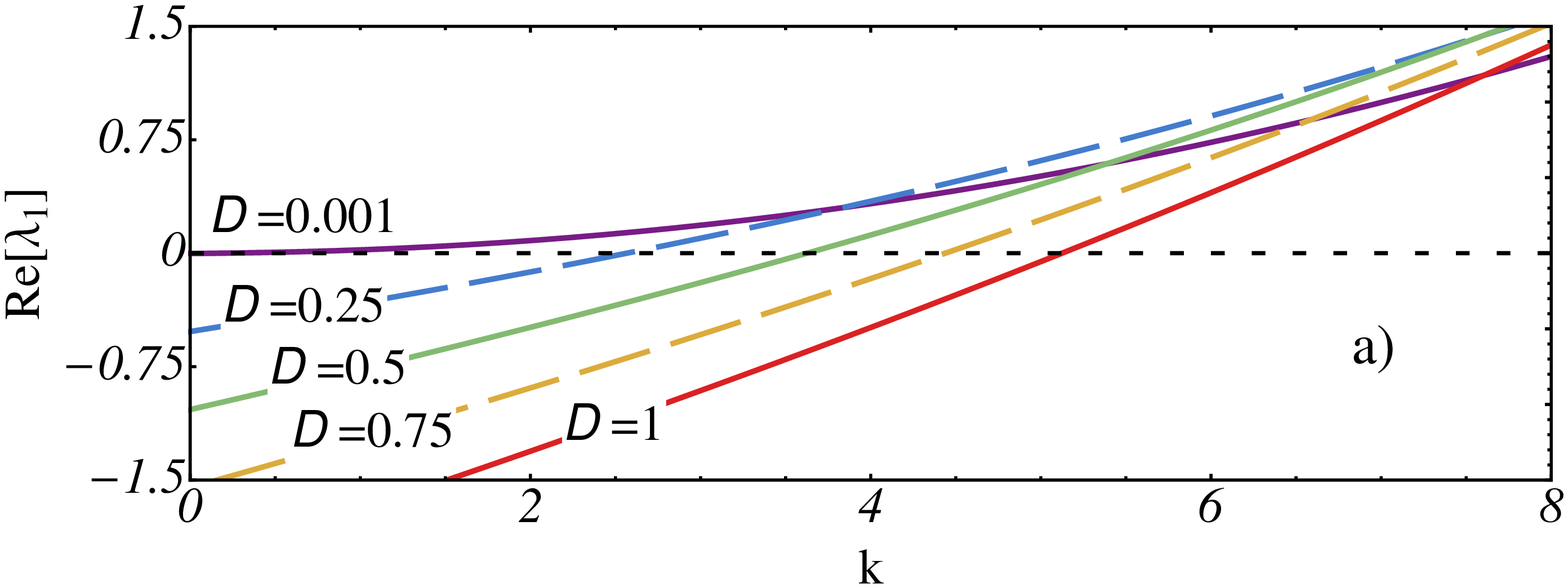}} {\includegraphics[width=75mm, height=30mm]{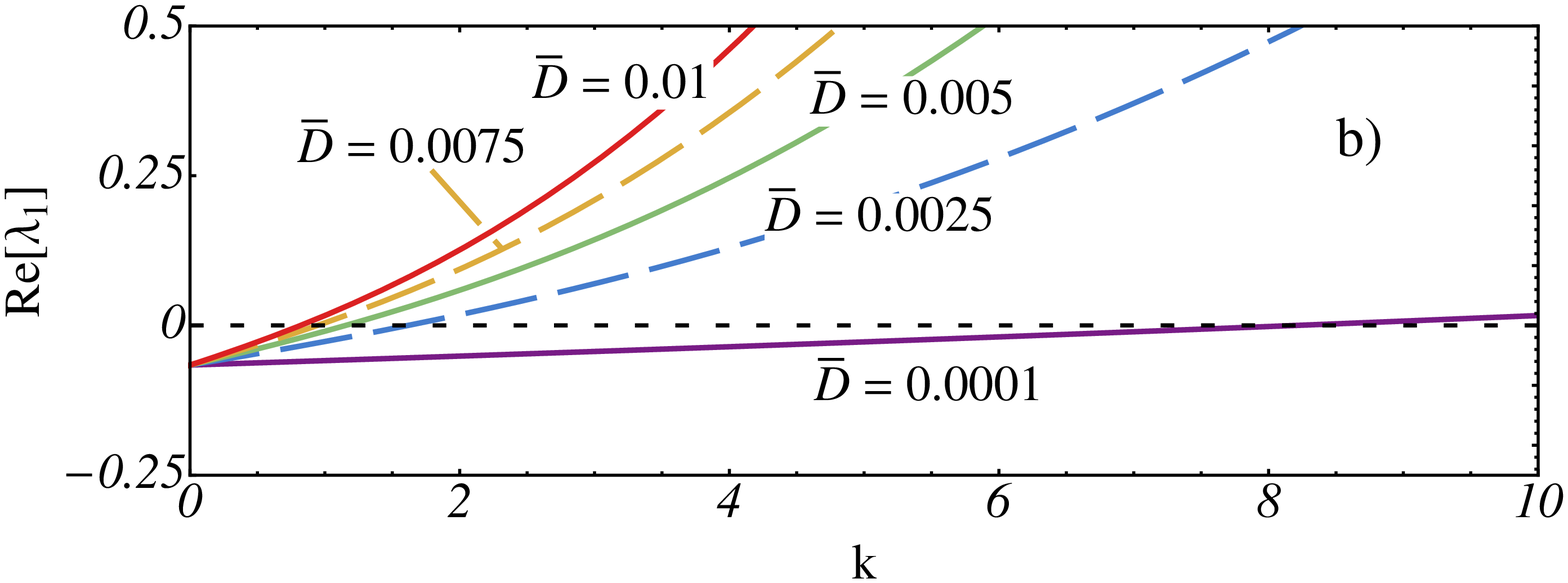}}
{\includegraphics[width=75mm, height=30mm]{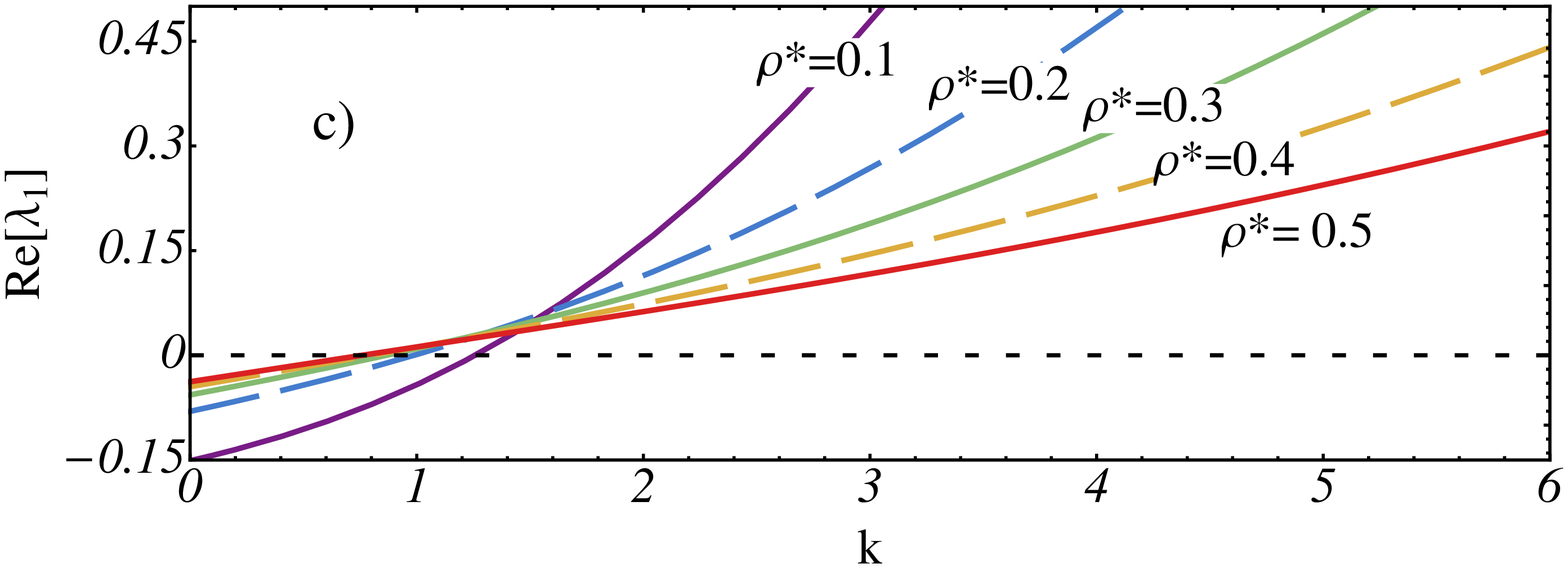}} {\includegraphics[width=75mm, height=30mm]{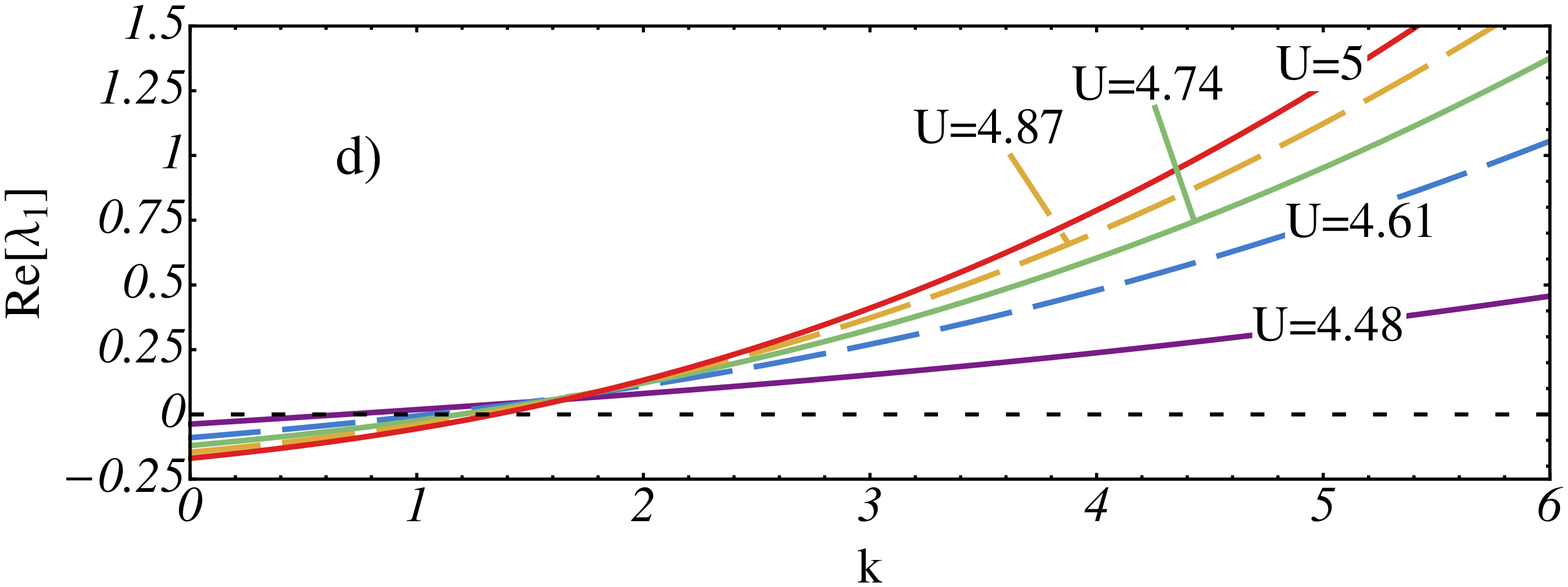}} 
\caption{The eigenvalue $\lambda_1$ of the linearized Eqs. (\ref{pattern03}) and (\ref{pattern04kroger}) 
as a function of the wave number $k$ in $cm^{-1}$ for different values of the parameters 
$\bar{D}$ in $cm^2\,s^{-1}$ (top); ${\cal D}$ in $s^{-1}$ and $\rho^*$ in $cm^{-3}$ (bottom). 
The eigenvalue becomes negative for $\rho^*$ larger than a certain critical value $\rho^*>\rho_c$.
This means that large densities prevent the transition and the appearance of patterns.}
\label{figure_003}
\end{figure}
\begin{eqnarray}
\frac{\partial S_{ij}}{\partial t} & = & -6 \mathcal{D} S_{ij} 
-2 \frac{ \mathcal{D}}{k_{B} T}  
\bigg\langle 
        \left[ \vec{u} \frac{\partial \mathcal{U}}
                            {\partial \vec{u}} 
        \right]^{s}_{ij}  
       - u_{i} u_{j} u_{k} \frac{\partial \mathcal{U}}{\partial u_{k} } 
\bigg\rangle \nonumber \\
& & + \bar{D} \nabla^{2} S_{ij} + D_{a} \nabla_{k} \nabla_{j} 
\bigg\langle
\left( u_{i} u_{j} -\frac{1}{3} \delta_{ij}  \right)
\left( u_{k} u_{l} -\frac{1}{3} \delta_{kl}  \right)
\bigg\rangle \nonumber \\ 
& & + \frac{\bar{D}}{k_{B} T} \nabla_{k} 
      \bigg\langle \left( u_{i} u_{j} -\frac{1}{3} \delta_{ij} \right)\nabla_{k} \mathcal{U} 
\bigg\rangle \nonumber \\
& & + \frac{D_{a}}{k_{B} T} \nabla_{k} 
\bigg\langle \left( u_{i} u_{j} -\frac{1}{3} \delta_{ij} \right)
        \left( u_{k} u_{l} -\frac{1}{3} \delta_{kl} \right)
        \nabla_{l} \mathcal{U}  
\bigg\rangle , \label{order001}
\end{eqnarray}
where the symbol $\partial / \partial \vec{u}$, represents the gradient
operator in $\vec{u}$-space. It is worth stressing that in the proper
limiting situations, Eq.~(\ref{order001}) reduces to diverse dynamic equations 
appearing in literature for the tensor order parameter.\cite{doi001,korgerbook,ilg001}
Eq. (\ref{order001}) is a particular case of an equation for polydomain nematic liquid 
crystals under shear stresses used to derive microscopic formulae for the 
Frank-Ericksen elastic coefficients.~\cite{kroeger001} 
As indicated previously, Eqs. (\ref{pattern03}) and (\ref{order001}) constitute a set of two coupled reaction-diffusion type partial differential
equations for the scalar fields $U\left( \vec{r} , t \right)$ and $S\left( \vec{r} , t \right)$.
\begin{figure}
{\includegraphics[width=75mm, height=30mm]{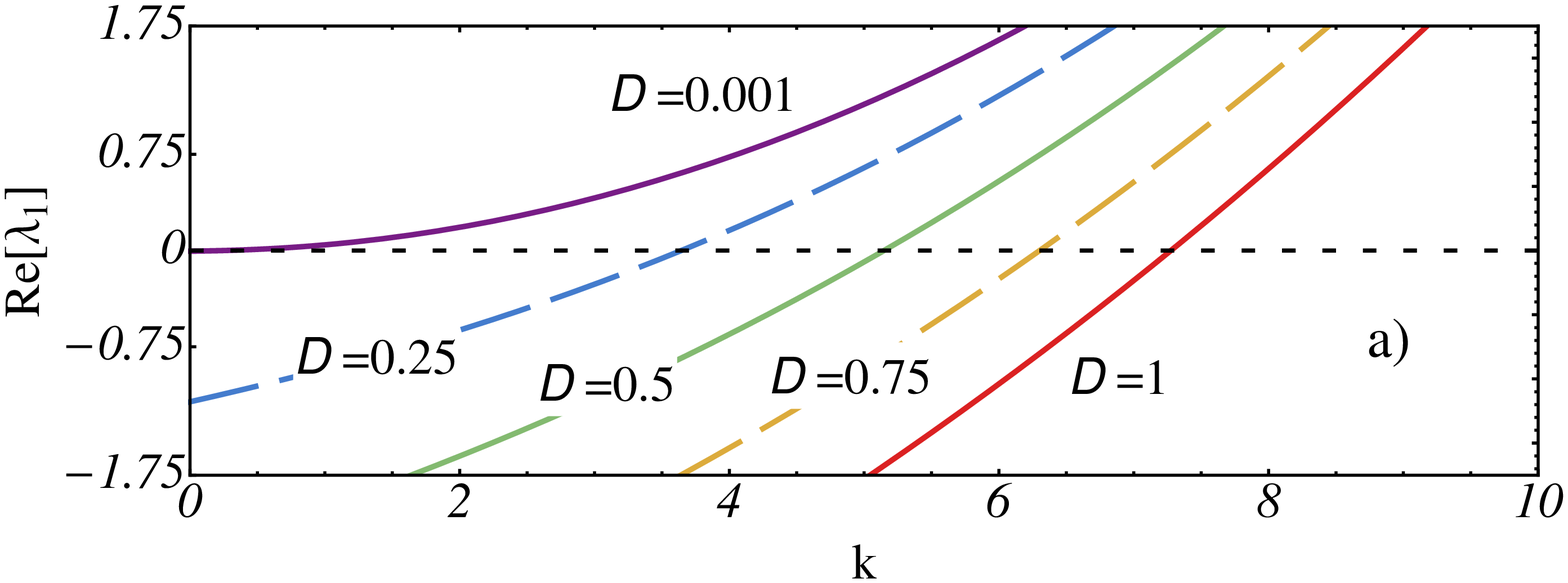}} {\includegraphics[width=75mm, height=30mm]{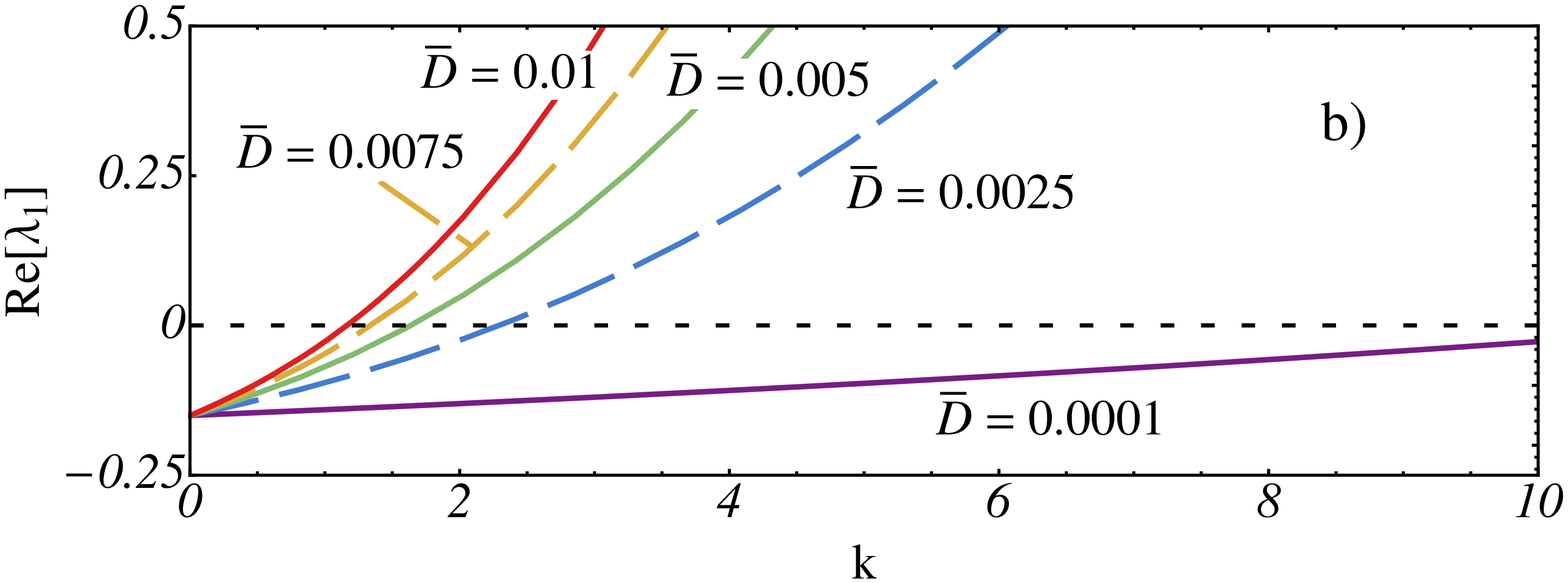}}
{\includegraphics[width=75mm, height=30mm]{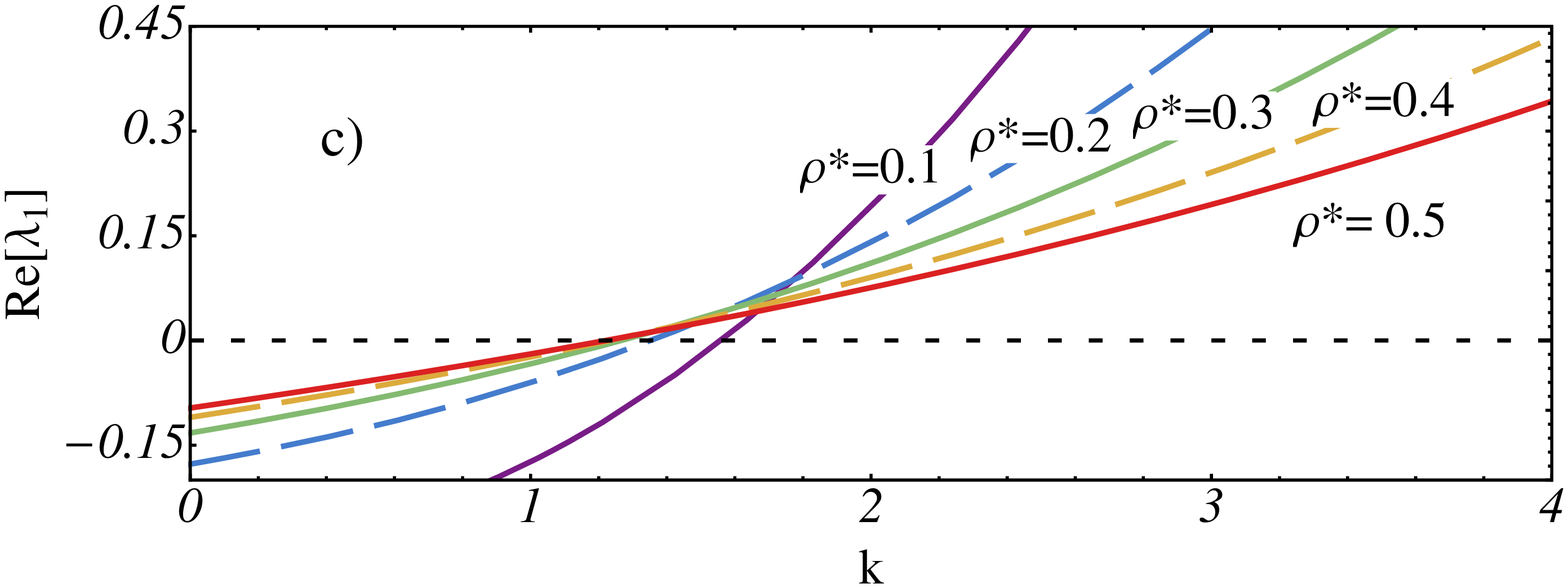}} {\includegraphics[width=75mm, height=30mm]{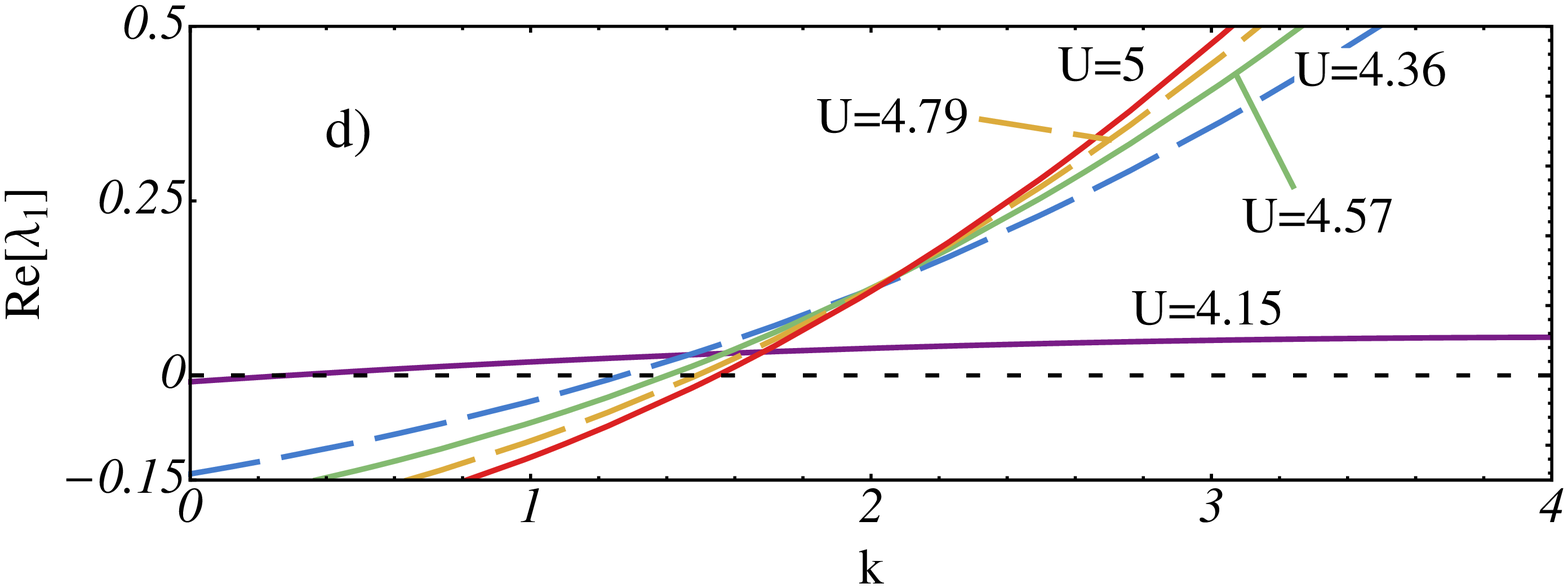}} 
\caption{The eigenvalue $\lambda_1$ of the linearized Eqs. (\ref{pattern03}) and (\ref{pattern04}) 
as a function of the wave number $k$ in $cm^{-1}$ for different values of the parameters 
$\bar{D}$ in $cm^2\,s^{-1}$ (top); ${\cal D}$ in $s^{-1}$ and $\rho^*$ in $cm^{-3}$ (bottom). 
As in the previous case, the eigenvalue may become negative for $\rho^*$ larger than a certain critical value $\rho^*>\rho_c$.}
\label{figure_004}
\end{figure}

With the aim to simplify the description, let us assume that $D_a =0$ and consider that the effective diffusion coefficient of the 
order parameter, $\bar{D}$, is again a scalar constant quantity. As a consequence of 
these assumptions and using Eqs. (\ref{closure00l}) and (\ref{closure00p}) the evolution equation for $S(\vec{r},t)$ reduces to
\begin{equation} 
\frac{\partial S}{\partial t}  = \bar{D} \nabla^2 S -6\mathcal{D} \left[(1-\frac{U}{5})S-\frac{1}{7} US^2 +\frac{12}{35} \nu U S^3 +\frac{6}{35}\nu(1-\nu)US^4\right],
\label{pattern04kroger}
\end{equation}
for the five order expansion of the free-energy like function in the parametric KAC closure,\cite{KAC} and
\begin{equation}
\frac{\partial S}{\partial t}  = \bar{D} \nabla^2 S -6\mathcal{D} \left[
            \left( 1 - \frac{U}{5} \right) S
            -\frac{U}{7} S^{2}
            + \frac{12}{35}U S^{\frac{13}{3}}  \right], 
\label{pattern04}
\end{equation}
for the dynamic closure. 

The existence of stationary patterns can be proved by analyzing Lyapunov's stability around the nematic
equilibrium state $S_{0}$ shown by the free-energy like functions (\ref{closure00l}) and (\ref{closure00p}).
In order to do this, we have to linearize the system of coupled equations (\ref{pattern03}) and (\ref{pattern04kroger})
(for the parametric closure),  and (\ref{pattern03}) and (\ref{pattern04}) (for the dynamic closure) about $S_{0}$ by 
assuming $S=S_{0}+\delta S$ and $U=U_{0}+\delta U$.\cite{prigo} 
This procedure allows us to calculate the elements of the matrix $\underline{\underline{\Lambda}}$ defining the resulting linear transformation
$\dot{\underline{X}} = \underline{\underline{\Lambda}}\cdot \underline{X}$, with $\underline{X} = (\delta S,\delta U)$ 
and $\dot{\underline{X}}$ the corresponding time derivative. The corresponding elements of these matrices are given in the Appendix A. 
\begin{figure}
\begin{center}
{\includegraphics[scale=.35]{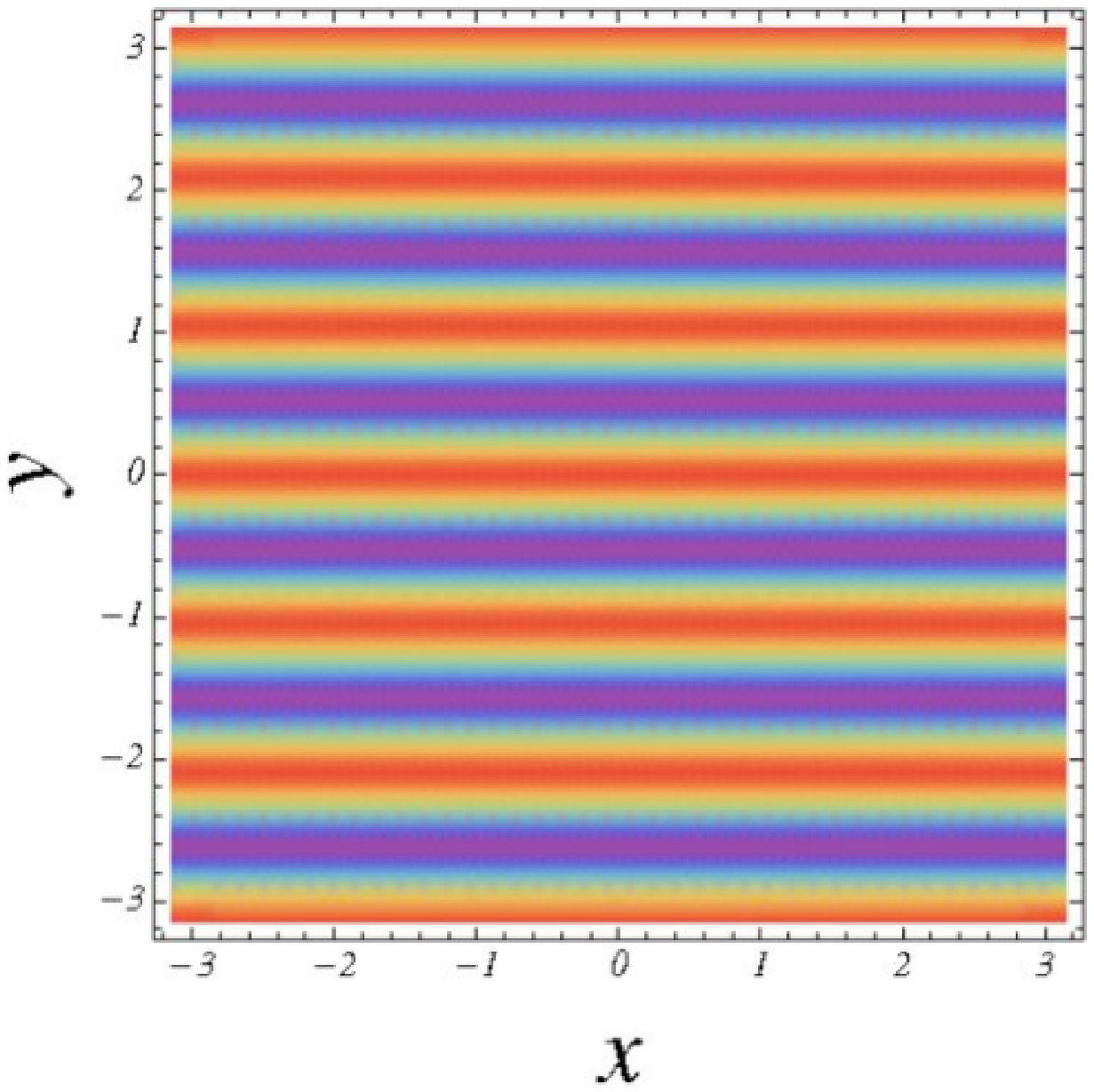}} 
{\includegraphics[scale=.35]{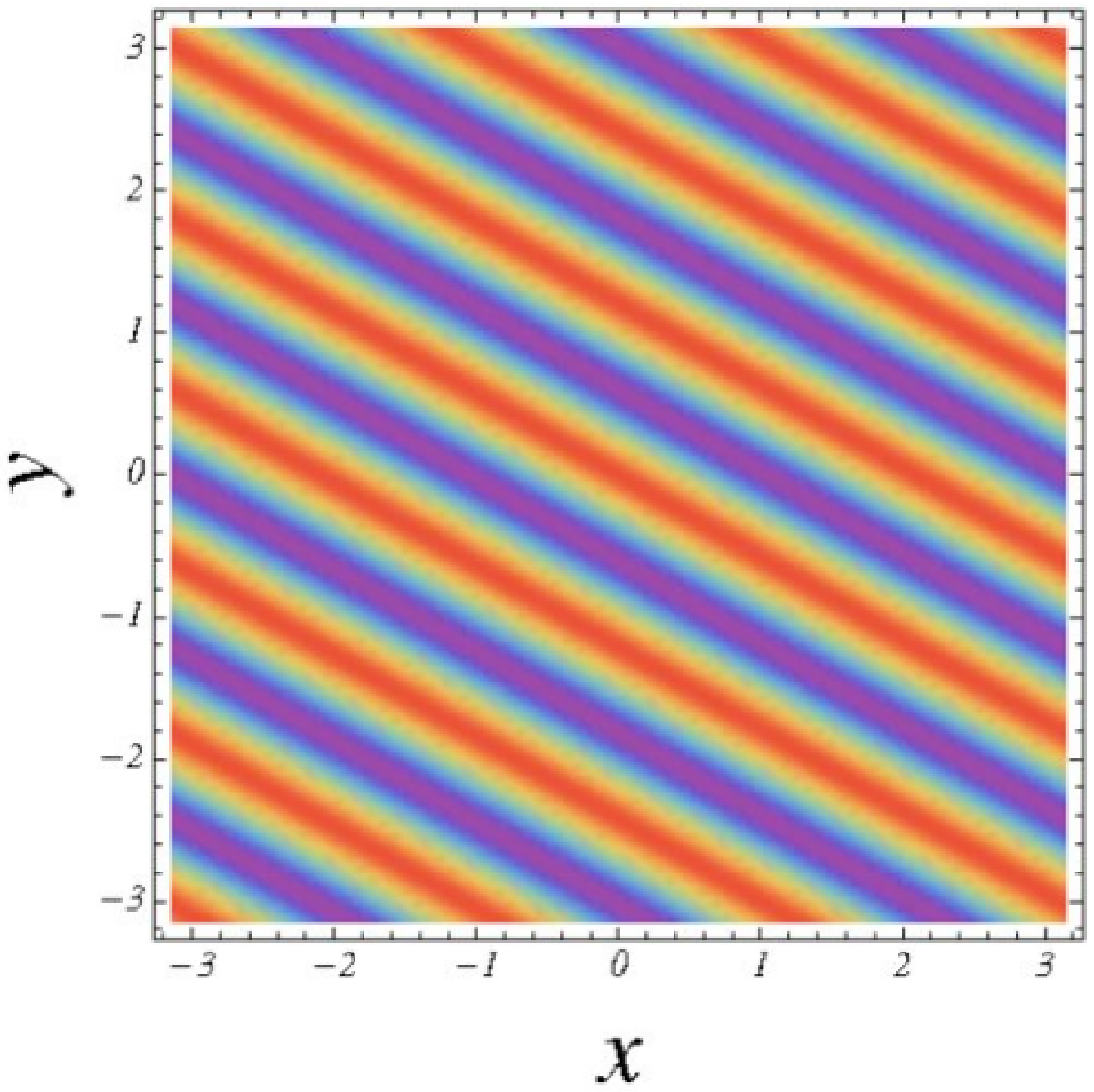}} 
{\includegraphics[scale=.35]{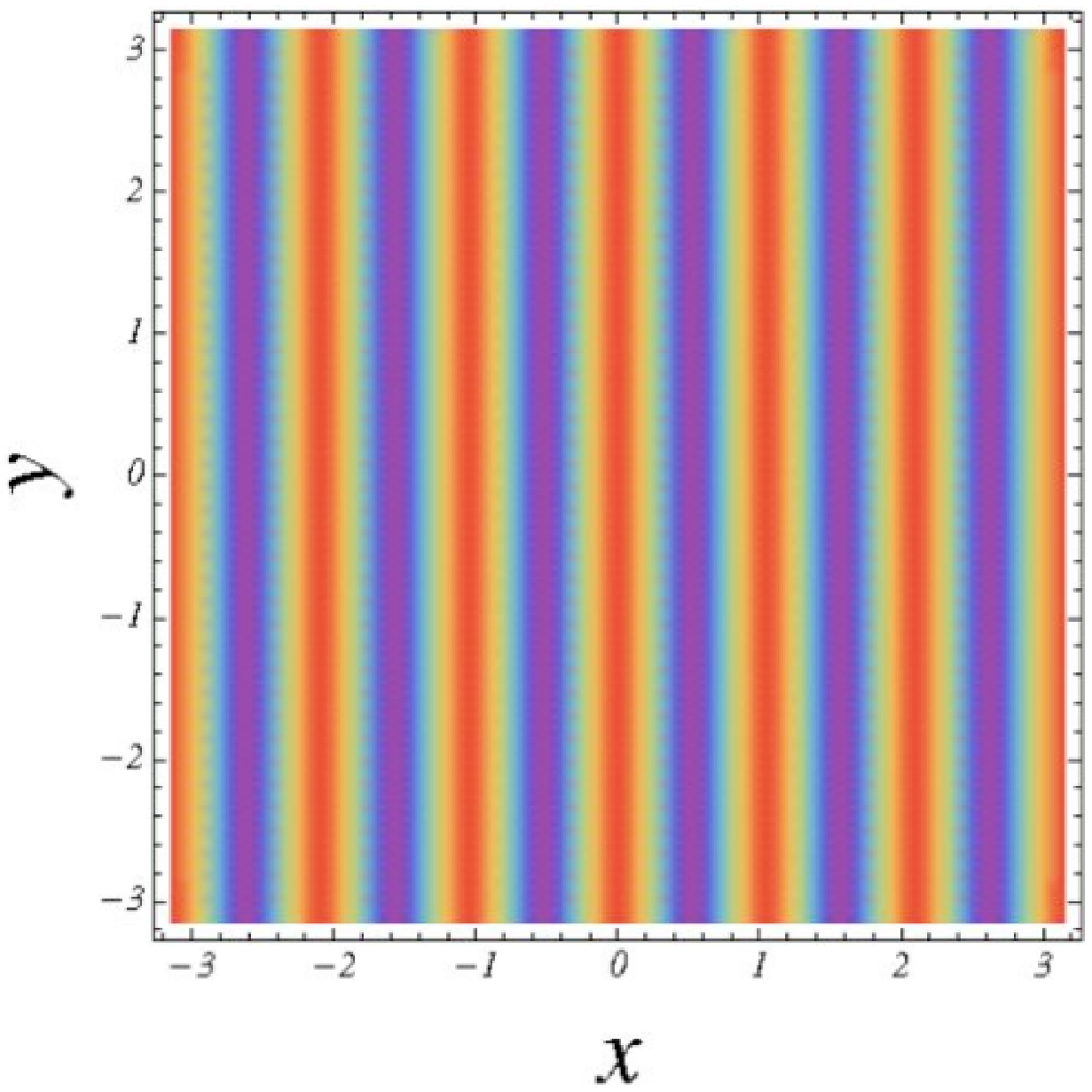}} 
\end{center}
\caption{Spatial structures of $\delta S$ for different components of a wave vector with magnitude $k=6\,cm^{-1}$. 
The colored regions correspond to different values of $S$ between $0$ and $1$. The dark purple regions 
correspond to the isotropic state ($S=0$) whereas the red regions 
correspond to a state with higher nematic order ($S=0.63$). The lighter regions separating the previous 
ones correspond to a nematic order between $0$ and $0.63$.}
\label{figure_005}
\end{figure}

The solution of the perturbative system $\dot{\underline{X}} = \underline{\underline{\Lambda}}\cdot \underline{X}$ 
can be proposed in terms of the combination
\begin{equation}\label{pattern05}
\underline{X}(\vec{r},t) = \sum_{\alpha} c_{\alpha} \underline{\Psi}_{\alpha} e^{i \vec{k}\cdot \vec{r}+\lambda_k t},
\end{equation}
where $\lambda_{\alpha}$ and $\underline{\Psi}_{\alpha}$ are the corresponding eigenvalues and eigenvectors of
$\underline{\underline{\Lambda}}$.

\begin{figure}
\begin{center}
{\includegraphics[scale=.3]{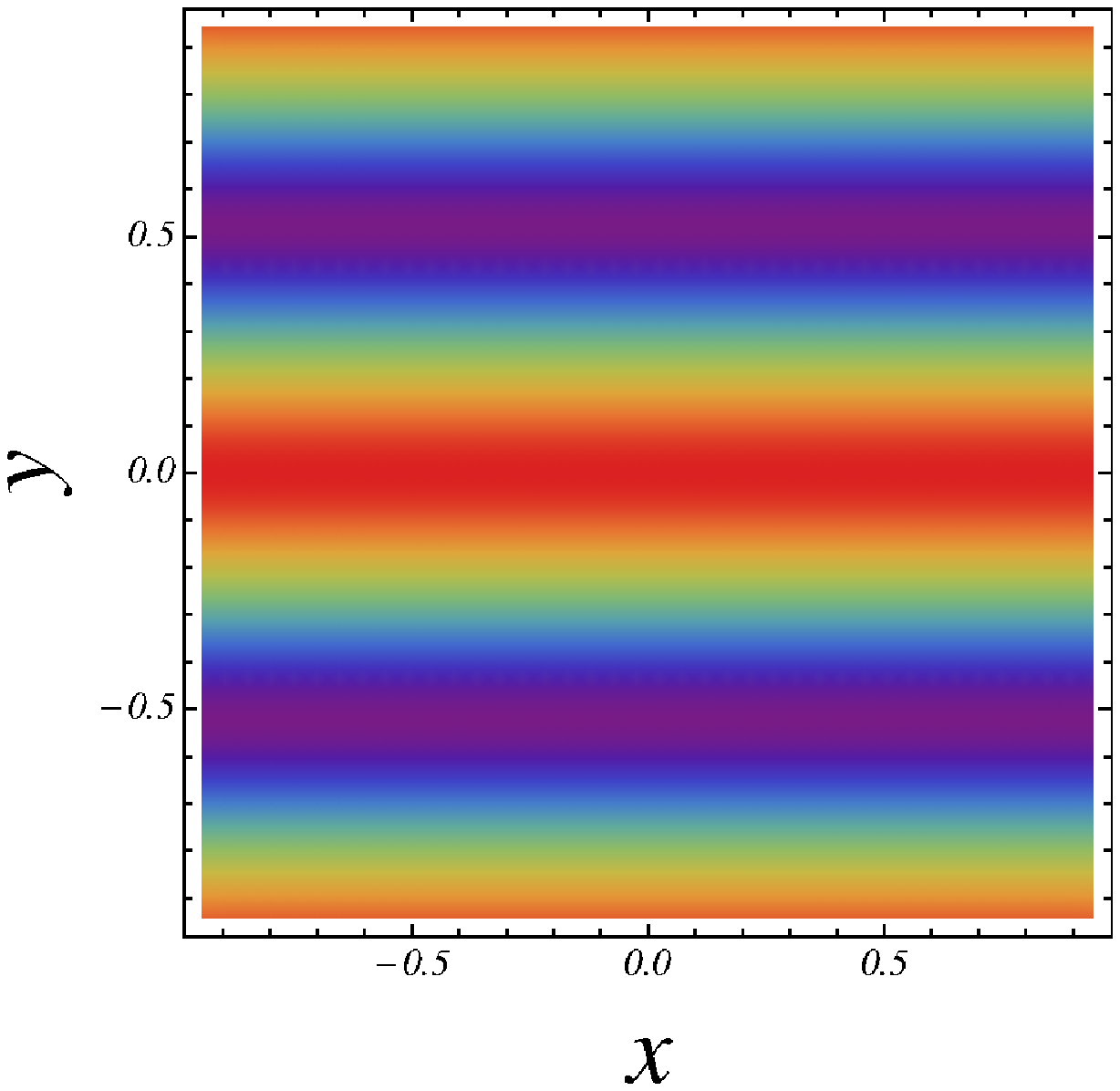}} {\includegraphics[scale=.3]{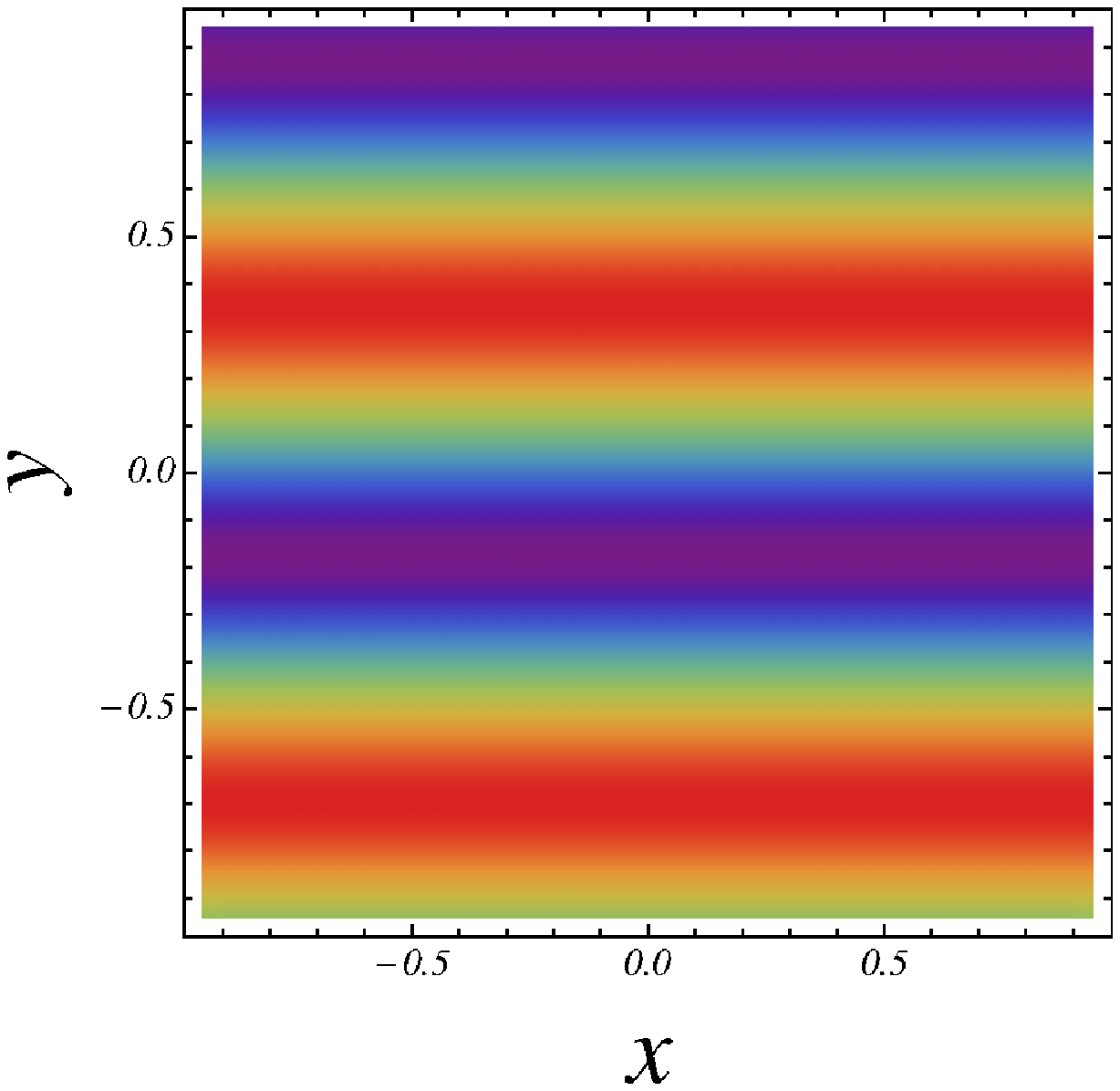}}
{\includegraphics[scale=.304]{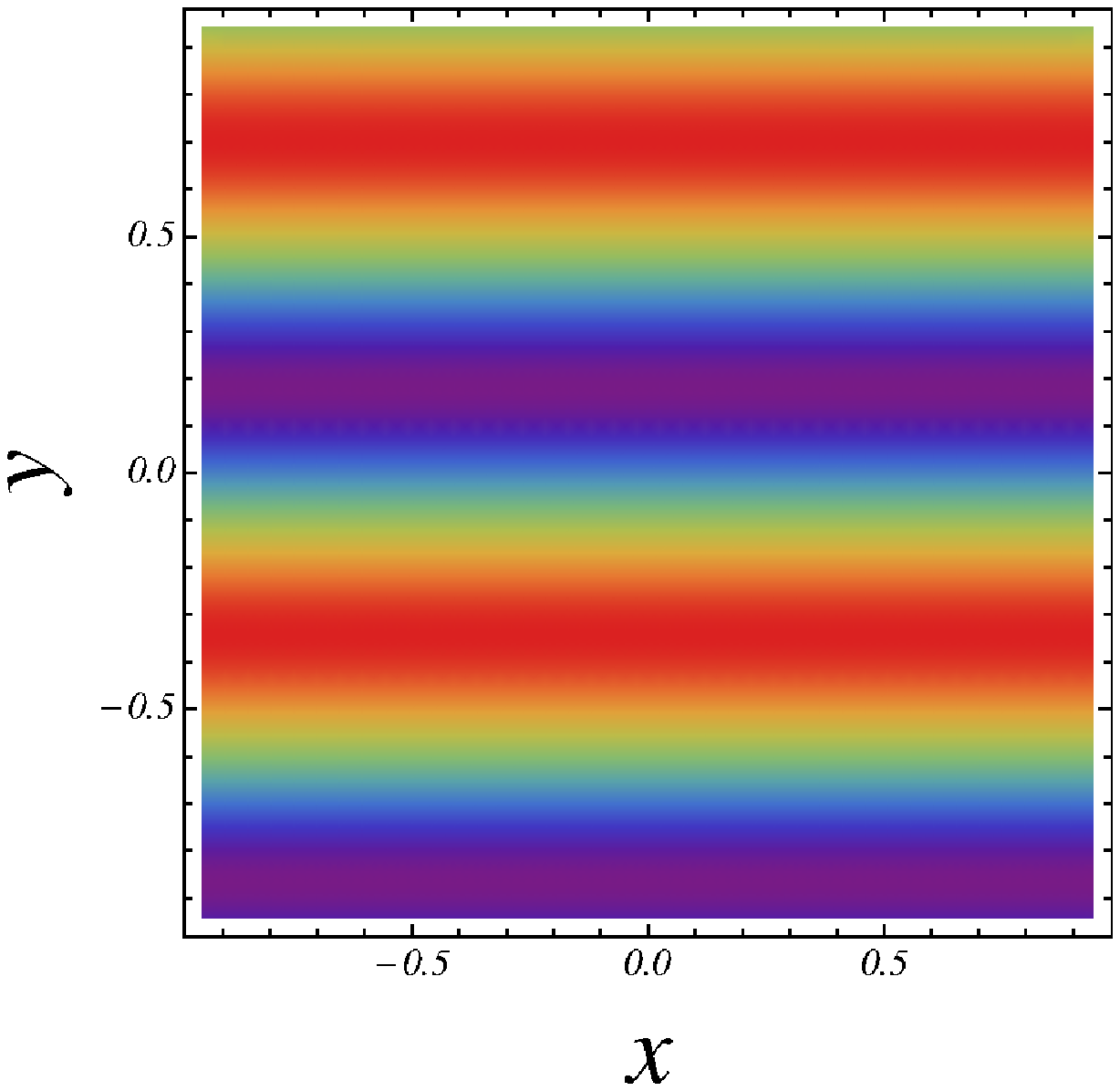}} {\includegraphics[scale=.306]{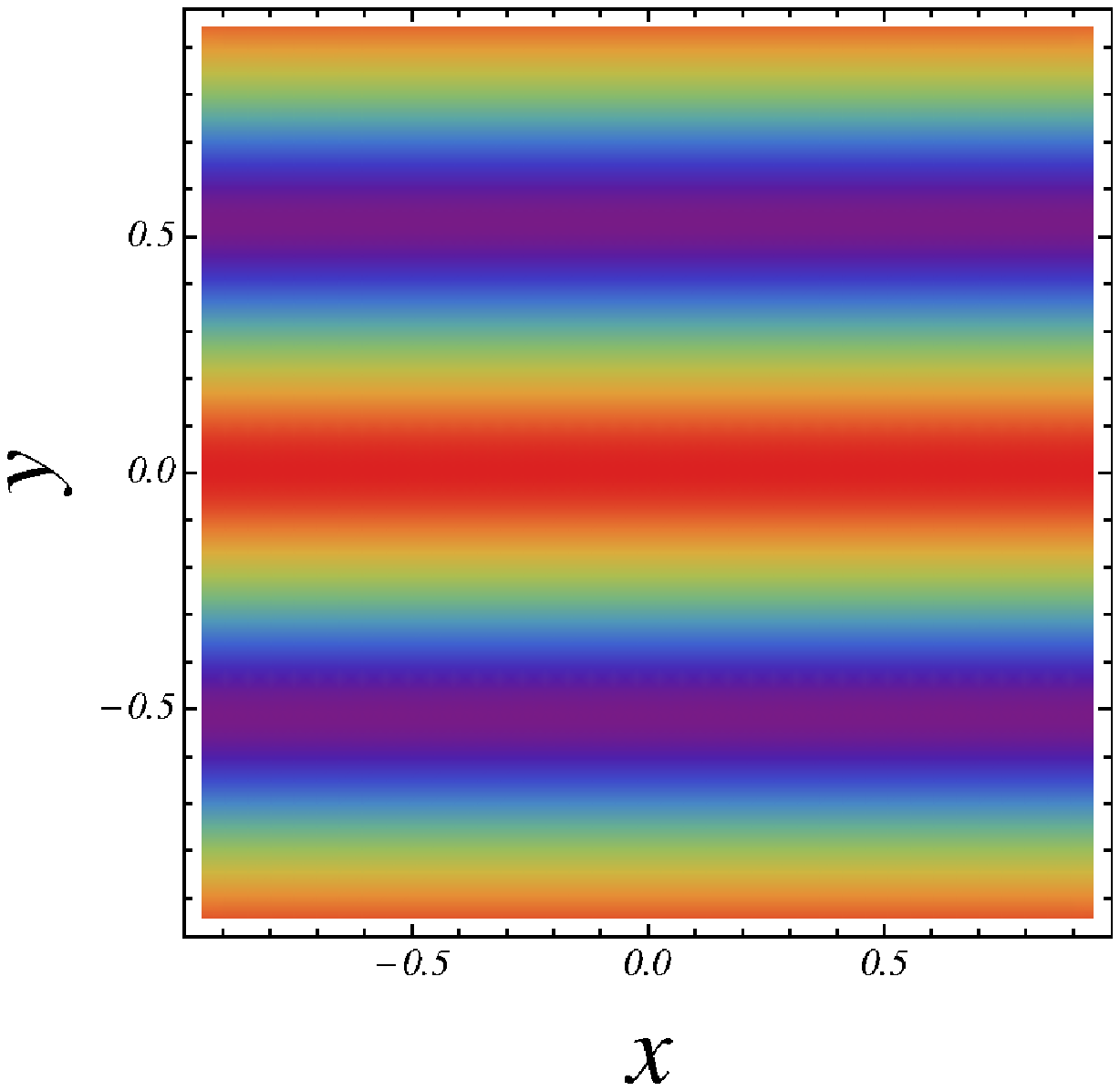}} 
\end{center}
\caption{ $\delta S$ as a propagating wave with $k=6\,cm^{-1}$. Each panel shows the configuration of the system for 
$t = 0, 4, 8, 12 s$. Colors mean the same as in Fig. 5.}
\label{figure_006}
\end{figure}

By performing a numerical study, it can be shown that both eigenvalues are complex. In both cases one of the eigenvalues
has a negative real part for all the combinations of values of the parameters $\bar{D}$, $\rho^*$ and $U_0$, whereas 
the other one may have a positive real part for certain combinations of the parameters. 
Figures~\ref{figure_003} and \ref{figure_004} show the corresponding results for the parametric and dynamic closures, respectively. 
These results imply that for the adequate combinations of parameters, i.e.
physical conditions, the system presents patterns that evolve in time in the form of traveling waves.\cite{prigo} 
The projections of these patterns on the $x-y$ plane are illustrated in Figure~\ref{figure_005} for 
a wave vector of magnitude $k=6\,cm^{-1}$ and three different combinations of its components. Figure~\ref{figure_006} illustrates 
the propagation of the patterns in time, for the same wave number.

Finally, we have also analyzed the case of the exact IHK equilibrium closure given in Ref. \cite{IHK} by assuming that the spatial and
temporal evolution of the scalar order parameter may be determined by an equation similar to Eqs. (\ref{pattern03}) and (\ref{pattern04kroger}),
and according to the phenomenological approach. The existence of patterns and traveling waves is also possible in that case.

\section{Discussion and conclusions}
In this article, we calculated a Fokker-Planck equation for the single-particle non-homogeneous distribution function of particle orientations 
by assuming an interaction energy of the Maier-Saupe type, which couples the mesoscopic degrees of 
freedom with the average value of the tensor order parameter $S_{ij}$. The FPE was derived by using the Gibbs entropy postulate
(\ref{mnet003}), that is also used for equilibrium situations, and calculating the entropy production of the system during relaxation. 
The obtained FPE is consistent with previous descriptions of the dynamics of liquid crystalline 
phases.~\cite{korgerbook,doi001,kroeger001}

Afterwards, we used the FPE to derive the first two coupled evolution equations for the moments of the distribution, 
Eqs.~(\ref{hierarchy006}) and (\ref{hierarchy007}) and truncated the corresponding hierarchy by adopting two dynamic closure approaches
in which the expression of the scalar fourth order parameter is expressed in terms of the scalar second order one.
The first parametric closure, is equivalent to the exact closure reported in Ref. \cite{KAC} whereas the second closure, called dynamic,
represents a novel closure that takes into account the leading term in the relaxation dynamics of the second and fourth scalar order parameters, 
Eqs. (\ref{closure00e}) and (\ref{closure00f}). This novel closure has a good performance when compared with the parametric closure
of Ref. \cite{KAC} and with the exact equilibrium closure of Ref. \cite{IHK}, as shown in Figures 1 and 2, and Table 1. 
A comparative discussion on the performance of the parametric and dynamic closures in the presence of shear flow was also done. 
From it may conclude that the dynamic closure relation, Eq.~(\ref{closure00n}), performs quantitatively very well when compared with the 
parametric closure Eq.~(\ref{closure00h}), and consequently can be also used 
to approximate the exact solution with high precision at least for the range of values considered here.

The non-equilibrium part of our analysis was motivated by the fact that Eqs. (\ref{closure00e}) and (\ref{closure00f})
are coupled in a similar way to Lotka-Volterra equations and, consequently, they
allow for the possibility of oscillating behaviors.\cite{KAC} We search for these behaviors in Section~5, 
where we derived two coupled partial differential equations of the diffusion-reaction type for the scalar second order parameter $S$
and the non-homogeneous degree of coupling $U$. A linear stability analysis of these coupled evolution equations showed that patterns and 
traveling waves are indeed possible non-equilibrium solutions for the closure approximations discussed in this paper 
and even for the case of the equilibrium closure of Ref. [19].

In summary, we studied the dynamics of uniaxial nematic systems by using the mesoscopic non-equilibrium thermodynamics
formalism in the context of a mean-field theory and within the approach of dynamic closures for the evolution equations for
the orientational order parameters. The results emerging from this analysis allows us to state that the non-equilibrium structures 
associated to pattern formation and traveling waves are possible for these systems without the influence of an external driving.

\appendix

\section{Explicit form of the linearized equations for $\delta S$ and $\delta U$}

As mentioned in Sec.V, the linearization of the coupled equations for $U$ and $S$ about $S_{0}$
by assuming $S=S_{0}+\delta S$ and $U=U_{0}+\delta U$ allows us to calculate the
elements of the matrix $\underline{\underline{\Lambda}}$ that enters in the linear transformation
$\dot{\underline{X}} = \underline{\underline{\Lambda}}\cdot \underline{X}$.
The elements of the corresponding Jacobian associated to the parametric closure, Eqs. (\ref{pattern03}) and (\ref{pattern04kroger}) are 
\small
\begin{eqnarray}\label{pattern05a}
\Lambda_{ss} = \bar{D} \nabla^2 
-\frac{ 6}{35}{\mathcal D} \left[\left(24 S_{0}^3 U_{0} \nu^2-24 S_{0}^3 U_{0} \nu-36 S_{0}^2 U_{0} \nu+10 S_{0} U_{0}+7
   U_{0}-35\right)\right] ,\\
\label{pattern05b}
\Lambda_{s\rho}=  -\frac{6}{35}{\mathcal D} \left[S_{0} \left(6 S_{0}^3 \nu^2-6 S_{0}^3 \nu-12 S_{0}^2 \nu+5 S_{0}+7\right)\right],  \\
\label{pattern05c}
 \Lambda_{\rho\rho}=-\frac{5 \bar{D}}{\rho^*} \left[3S_{0}\nabla U_{0}\cdot\nabla+3\nabla S_{0}\cdot\nabla U_{0}+2S_{0}\nabla^2 U_{0}+2U_{0}\nabla S_{0}\cdot\nabla+U_{0} S_{0}\nabla^2 + U_{0}\nabla^2 S_{0}\right]  , \\
\label{pattern05d}
\Lambda_{\rho s}=\bar{D}\nabla^2 -\frac{5 \bar{D}}{\rho^*} \left[3S_{0}\nabla S_{0}\cdot\nabla+S_{0}^2\nabla^2 +\nabla S_{0}\cdot\nabla S_{0}+S_{0}\nabla^2 S_{0}\right]  .
\end{eqnarray}
\normalsize
In order to search for patterns, we have to propose solutions of the form $\underline{X} =\underline{X}(t) e^{i \vec{k}\cdot \vec{r}}$ 
with $\vec{k}$ the wave vector of the perturbation. Assuming that in Eqs.  (\ref{pattern05a})-(\ref{pattern05d}) we may approximate
$\nabla S_0 \sim S_0/{\cal L}$ and $\nabla S_0 \cdot \nabla \sim \left(S_0/{\cal L}\right) ik$, we finally obtain the Fourier transformed relations
\small
\begin{eqnarray}\label{pattern05aa}
 \Lambda_{ss} = - k^2 \bar{D}  
-\frac{ 6}{35}{\mathcal D} \left[\left(24 S_{0}^3 U_{0} \nu^2-24 S_{0}^3 U_{0} \nu-36 S_{0}^2 U_{0} \nu+10 S_{0} U_{0}+7
   U_{0}-35\right)\right],      \\
\label{pattern05bb}
\Lambda_{s\rho}= -\frac{6}{35}{\mathcal D} \left[S_{0} \left(6 S_{0}^3 \nu^2-6 S_{0}^3 \nu-12 S_{0}^2 \nu+5 S_{0}+7\right)\right],  \\
\label{pattern05cc}
 \Lambda_{\rho\rho}= -\frac{5 \bar{D}}{\rho^*} S_{0} U_{0}\left[5\frac{ik}{\cal{L}}+\frac{6}{\mathcal{L}^2}-k^2\right],   \\
\label{pattern05dd}
 \Lambda_{\rho s}= -k^2 \bar{D}-\frac{5 \bar{D}}{\rho^*}S_{0}^2\left[3\frac{ik}{\mathcal{L}}+\frac{2}{\mathcal{L}^2}- k^2\right]. 
\end{eqnarray}
\normalsize
where ${\cal L} = \left[\bar{D}/{\cal D}\right]^{1/2}$ is a characteristic length of the system.

In similar form, the elements of the corresponding Jacobian associated to the dynamic closure, Eqs. (\ref{pattern03}) and (\ref{pattern04}) are 
\small
\begin{eqnarray}\label{pattern05aaa}
 \Lambda_{ss} = \bar{D} \nabla^2 
-\frac{ 6}{35}{\mathcal D} \left[\left(52 S_{0}^{10/3}-10 S_{0}-7\right) U_{0}+35\right] ,        \\
\label{pattern05bbb}
\Lambda_{s\rho}=  -\frac{6}{35}{\mathcal D} \left[S_{0} \left(12 S_{0}^{10/3}-5 S_{0}-7\right)\right],  \\
\label{pattern05ccc}
  \Lambda_{\rho\rho}=-\frac{5 \bar{D}}{\rho^*} \left[3S_{0}\nabla U_{0}\cdot\nabla+3\nabla S_{0}\cdot\nabla U_{0}+2S_{0}\nabla^2 U_{0}+2U_{0}\nabla S_{0}\cdot\nabla+U_{0} S_{0}\nabla^2 + U_{0}\nabla^2 S_{0}\right]  ,  \\
\label{pattern05ddd}
 \Lambda_{\rho s}=\bar{D}\nabla^2 -\frac{5 \bar{D}}{\rho^*} \left[3S_{0}\nabla S_{0}\cdot\nabla+S_{0}^2\nabla^2 +\nabla S_{0}\cdot\nabla S_{0}+S_{0}\nabla^2 S_{0}\right]  .
\end{eqnarray}
\normalsize
The corresponding Fourier transformed relations are in this case
\small
\begin{eqnarray}\label{pattern05aaaa}
\Lambda_{ss} = - k^2 \bar{D}  
-\frac{ 6}{35}{\mathcal D} \left[\left(52 S_{0}^{10/3}-10 S_{0}-7\right) U_{0}+35\right],  \\
\label{pattern05bbbb}
\Lambda_{s\rho}= -\frac{6}{35}{\mathcal D} \left[S_{0} \left(12 S_{0}^{10/3}-5 S_{0}-7\right)\right],  \\
\label{pattern05cccc}
 \Lambda_{\rho\rho}= -\frac{5 \bar{D}}{\rho^*} S_{0} U_{0}\left[5\frac{ik}{\cal{L}}+\frac{6}{\mathcal{L}^2}-k^2\right],     \\
\label{pattern05dddd}
 \Lambda_{\rho s}= -k^2 \bar{D}-\frac{5 \bar{D}}{\rho^*}S_{0}^2\left[3\frac{ik}{\mathcal{L}}+\frac{2}{\mathcal{L}^2}- k^2\right].   
\end{eqnarray}
\normalsize

\section*{Acknowledgments} 
We acknowledge Prof. M. Rubi by critically reading this manuscript. HH acknowledges the postdoctoral program by UNAM-DGAPA for 
financial support.  ISH and DMH thank UNAM-DGAPA for partial financial support of Grants No. IN102609 and ID100112-2.

\end{document}